\documentclass[aps,prb,twocolumn,showpacs,superscriptaddress]{revtex4-1}
\usepackage{dsfont,amsthm,amsbsy}
\usepackage{verbatim}
\usepackage{amssymb}
\usepackage{amsmath}
\usepackage{bbm}
\usepackage{graphicx}
\usepackage{epstopdf}
\usepackage{subfigure}
\usepackage{natbib}
\usepackage{epsfig}
\usepackage{amsfonts}
\usepackage{mathrsfs}
\usepackage{sidecap}
\usepackage{lipsum}
\usepackage[toc,page,title,titletoc,header]{appendix}
\usepackage[colorlinks,linkcolor=blue,citecolor=blue,anchorcolor=blue,urlcolor=blue]{hyperref}
\usepackage{hyperref}
\usepackage{resizegather}
\usepackage{tikz}
\usepackage{float}
\usepackage{mathbbol}
\usepackage[normalem]{ulem}
\usepackage{cancel}
\usepackage{upgreek}

\newcommand{\bl}{\begin{aligned}}
\newcommand{\el}{\end{aligned}}
\def\be{\begin{equation}}
\def\ee{\end{equation}}

\def\bi{\begin{itemize}}
\def\ei{\end{itemize}}
\def\bn{\begin{enumerate}}
\def\en{\end{enumerate}}
\def\bea{\begin{eqnarray}}
\def\eea{\end{eqnarray}}
\def\no{\nonumber}
\def\ba{\begin{array}}
\def\ea{\end{array}}
\def\bd{\begin{displaymath}}
\def\ed{\end{displaymath}}

\begin{document}

\title{Dynamical Topological Quantum Phase Transitions at Criticality}

\author{M. Sadrzadeh}
\email{marzieh.sadrzadeh@gmail.com}
\affiliation{Department of Physics, Institute for Advanced Studies in Basic Sciences (IASBS), Zanjan 45137-66731, Iran}
\author{R. Jafari}
\email[]{jafari@iasbs.ac.ir, rohollah.jafari@gmail.com}
\affiliation{Department of Physics, Institute for Advanced Studies in Basic Sciences (IASBS), Zanjan 45137-66731, Iran}
\affiliation{Department of Physics, University of Gothenburg, SE 412 96 Gothenburg, Sweden}
\affiliation{Beijing Computational Science Research Center, Beijing 100094, China}
\author{A. Langari}
\email[]{langari@sharif.edu}
\affiliation{Department of Physics, Sharif University of Technology, P.O.Box 11155-9161, Tehran, Iran}
%\date{\today}

\begin{abstract}
The nonequilibrium dynamics of two dimensional Su-Schrieffer-Heeger model, in the presence of staggered chemical potential,
is investigated using the notion of dynamical quantum phase transition.
We contribute to expanding the systematic understanding of the interrelation between the equilibrium quantum phase transition and the dynamical quantum phase transition (DQPT).
Specifically, we find that dynamical quantum phase transition relies on the existence of massless {\it propagating quasiparticles} as signaled by their impact on the Loschmidt overlap. These massless excitations are a subset of
all gapless modes, which leads to quantum phase transitions. The underlying two dimensional model reveals gapless
modes, which do not couple to the dynamical quantum phase transitions, while relevant massless quasiparticles
present periodic nonanalytic signatures on the Loschmidt amplitude. The topological nature of  DQPT is verified by the quantized integer values of the topological order parameter, which gets even values.
Moreover, we have shown that the dynamical topolocical order parameter truly captures the topological phase transition on the zero Berry curvature line, where the Chern number is zero and the two dimensional Zak phase is not the proper idicator.
\end{abstract}


\maketitle

\section{introduction}
Quenching a quantum system continues to pose a challenging problem in several disciplines of current research activities \cite{Polkovnikov2011,Mitra2018,Dutta:2017717}.
The unitary time evolution of a quantum system after a sudden global quench
plays substantial role in understanding non-equilibrium quantum physics \cite{Quan2006,Jafari2017,Essler2016,Campbell2016,Mishra2016,Fogarty2020,Bayat2018}.
The sudden quantum quench process can be experimentally implemented
\cite{Jurcevic2017,Vogel2017,Guo2019,Tian2019,Wang2019}
and being formulated in theoretical calculations \cite{Heyl2014,Vajna2015,Heyl2018,Abdi2019,Yang2019,Sedlmayr2018,Sedlmayr2018b,Jafari2019,Srivastav2019,Bhattacharya2017,
Bhattacharya2017b,Huang2016,Lahiri2019,Budich2016,Jafari2016,Mishra2018,Mendl2019,Mishra2020,Zache},
which attracts attentions to reveal its unknown aspects.
Accordingly, a new area of research named {\it dynamical quantum phase transitions}
has been turned out \cite{Heyl2013} in non-equilibrium quantum and statistical physics.

DQPT proposed as a counterpart of conventional equilibrium thermal phase transitions,
where real time becomes analogous to a control parameter such as temperature.
Whithin an equilibrium phase transition the free energy corresponding to an equilibrium partition function
becomes nonanalytic as a function of a control parameter, while DQPT is signalled by real-time nonanalyticities in the rate
function of the Loschmidt overlap (LO), i.e. the overlap between initial and its time evolved wave function \cite{Heyl2013,Quan2006}.
These nonanalyticities are accompanied by zeros of Loschmidt overlap, which are known in statistical physics as Fisher zeros of the partition function  \cite{Yang1952,Fisher1978,Fisherbook,Heyl2013,Lang}.
Initial theoretical findings showed that DQPT and the equilibrium quantum phase transition (EQPT) are ultimately related: nonanalyticities in the rate function of LO occur only for quenches crossing the EQPT point \cite{Heyl2013}. This corespondence was verified in several subsequent studies for both non-integrable \cite{Karrasch2013,Kriel2014,Canovi2014,Palmai2015,Halimeh2017,Hubig2019} and integrable \cite{Schmitt2015,Bhattacharya2017c,Dutta2017,Bhattacharjee2018,Zamani2020,Jafari2020a,Porta2020,Puebla} models.
Further studies, however, reveal that DQPT is not in a one-to-one correspondence to EQPT, \cite{jafari2019dynamical,Vajna2014,Andraschko2014,Sharma2015,Uhrich2020,Sedlmayr2020,Khatun2019} suggesting that DQPT can, in general, be rather seen as a critical phenomenon distinct from the physics of EQPT \cite{Heyl2018,Uhrich2020,ding2020dynamical}.

\iffalse
Remarkably, analogous to order parameters at conventional EQPT, it was established using the Pancharatnam geometric phase \cite{Samuel1988,Budich2016}, that there exists a dynamical topological order parameter (DTOP) which can characterize DQPTs \cite{Budich2016}. The presence of a DTOP indicates the emergence of a topological feature associated with the time evolution of quenched systems. The DTOP takes integer values as a function of time and displays quantized jumps at the critical times, signaling the existence of DQPTs \cite{Budich2016,Bhattacharya2017c,Bhattacharjee2018}.
{\color{red}Recently the DQPT has been studied in transverse field XY model \cite{ding2020dynamical} for quench from/to critical point and for a quench along the critical line. It has been shown that, the system shows DQPT and the dynamical topological order parameter is half-quantized or unquantized.}
\fi

%
%%%%%%%%%%%%%%%%%%%%%%%  Fig.1   %%%%%%%%%%%%%%%%%%%%%%%
\begin{figure*}[t!]
\centerline{
\includegraphics[width=0.244\linewidth]{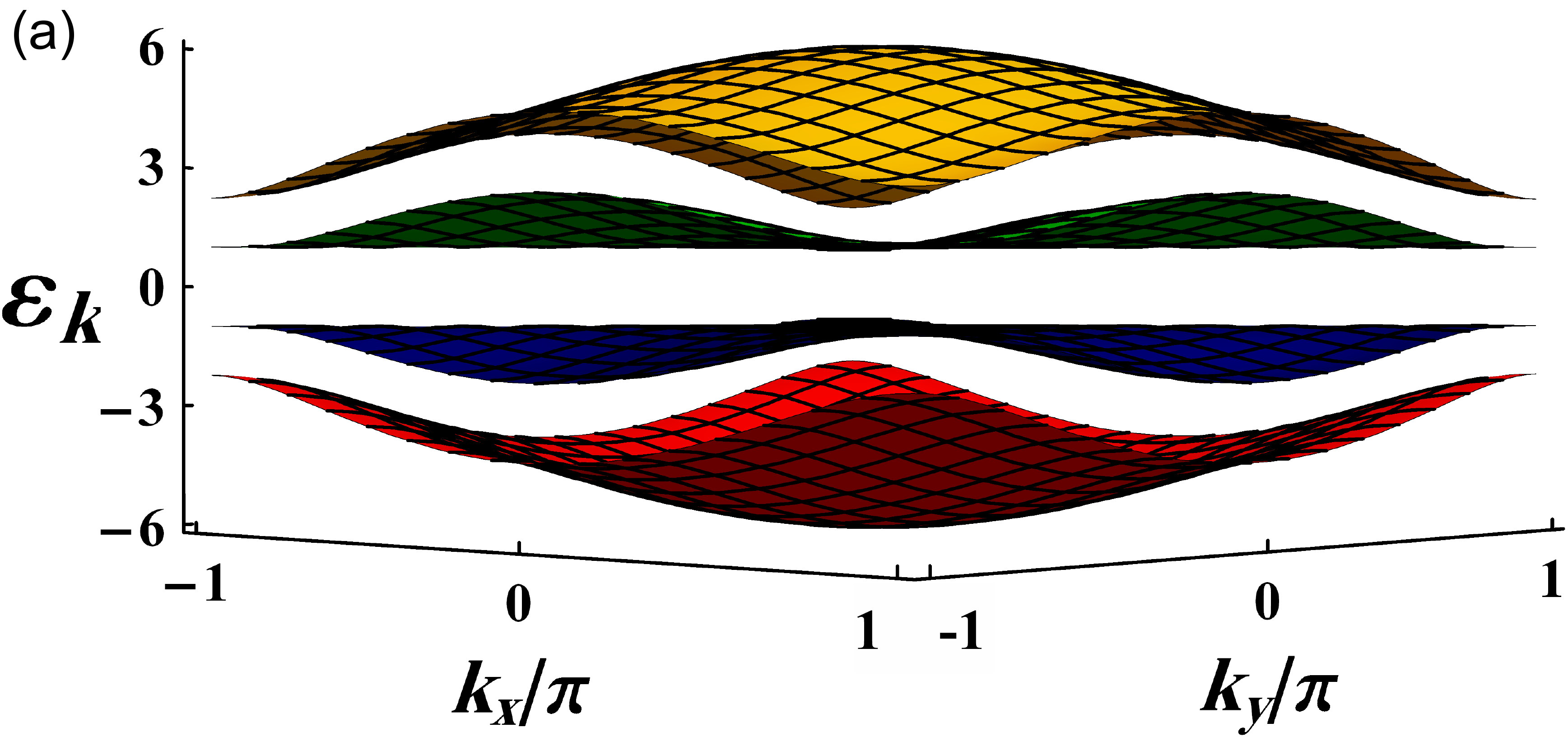}
\includegraphics[width=0.244\linewidth]{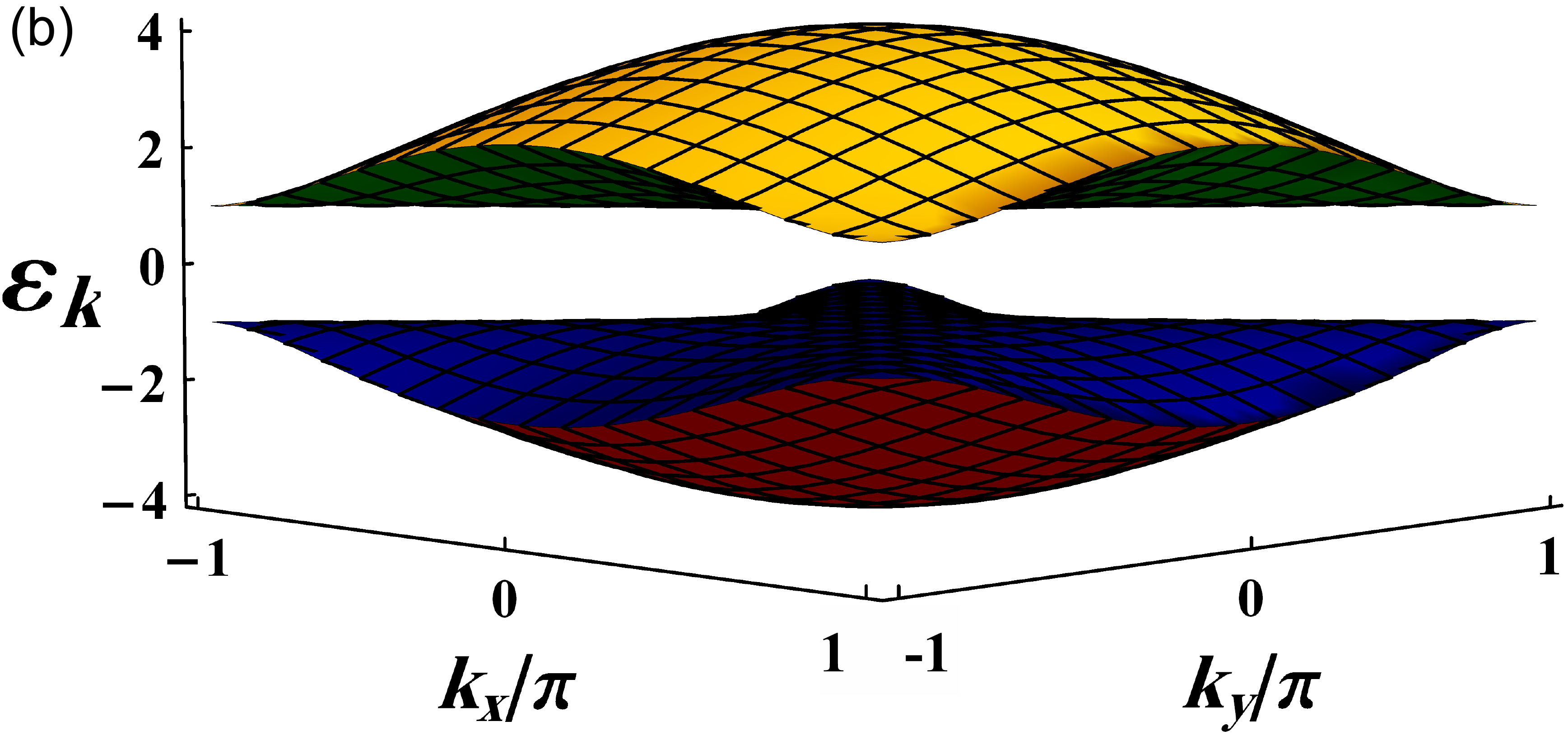}
\includegraphics[width=0.244\linewidth]{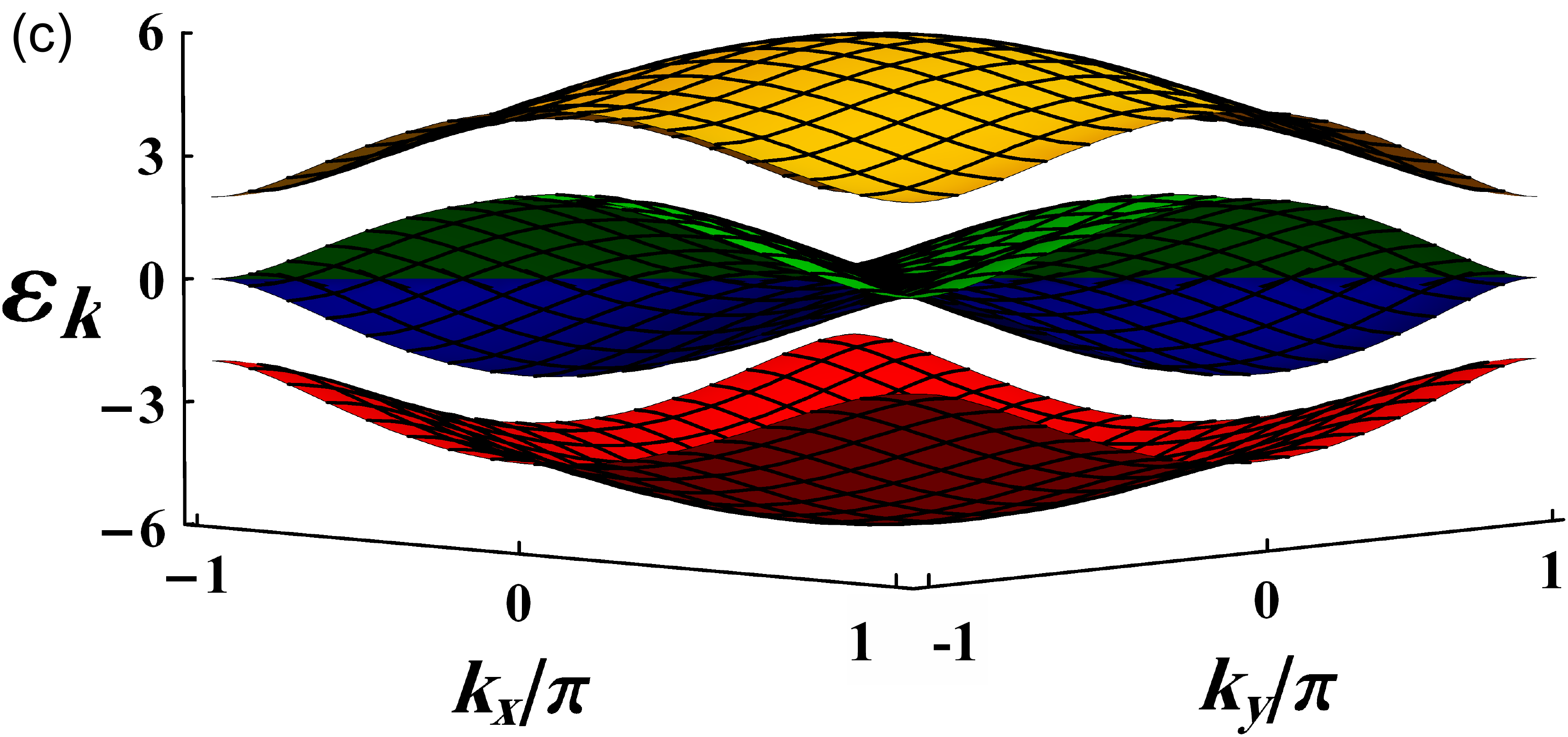}
\includegraphics[width=0.268\linewidth]{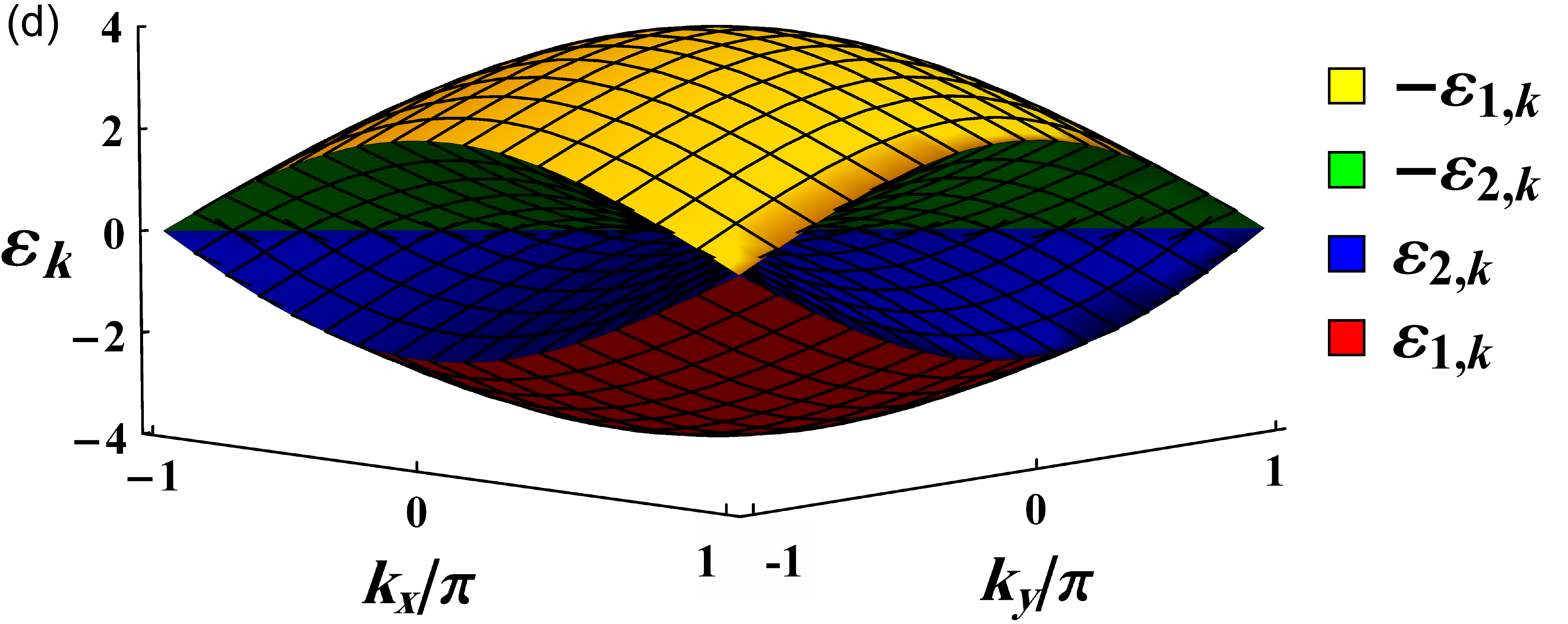}}
\caption{(Color online) Energy spectrum of the extended 2D SSH model for nonzero chemical potential $\mu=1$ at (a) anisotropic point, $w=1$, $v=2$ and (b) isotropic point $w=v=1$. The spectrum for zero chemical potential $\mu=0$, at (c) anisotropic point $w=1$, $v=2$ and (d) isotropic point $w=v=1$.}
\label{fig1}
\end{figure*}
%%%%%%%%%%%%%%%%%%%%%%%%%%%%%%%%%%%%%%%%%%%%%%%%%%%%%%%
%

In this paper, we revisit the DQPT to highlight the conditions under which DQPT can be observed in a
quantum system.
To this aim, by going beyond the two bands models \cite{ding2020dynamical}, we investigate the dynamics of the two dimensional (2D) exactly solvable Su-Schrieffer-Heeger (SSH) model having alternating hopping amplitudes ($w$, $v$) in both directions on the square
lattice in the presence of a staggered chemical potential ($\mu$). We show that, the zero chemical potential
leads to a highly degenerate gapless critical line, for any arbitrary ratio of the alternating hopping amplitudes. The model is always gapped for nonzero chemical potentials.
We study the dynamical phase transition in the whole phase diagram of the model parameters,
namely ($\frac{\mu}{w}, \frac{v}{w}$).
We do not observe DQPT along the critical (gapless) line except for quenching across the topological quantum phase transition (TQPT) point. The topological phase transition-- from trivial to non-trivial topological phases-- occurs at a single critical point, where the hopping amplitudes are equal. In other words, we illustrate that quenching across the EQPT point is neither a necessary nor a sufficient condition for the presence of DQPT.
Instead, the quasiparticle gapless modes, which control the Loschmidt overlap determine the appearance of DQPT.
If the latter modes coincide with the equilibrium quantum critical counterparts, the periodical nonanalytic structure of the rate function can tie to the EQPT. This condition, which is general, shed new light on the important issue of how to interpret the LO after a quantum quench. Moreover, the Zak phase, which always indicates topological phase
transitions in one-dimensional models, is not a reliable signature for 2D topological phase transitions.

The paper is organized as follows. In Sec.~\ref{Model}, we introduce the model
and the method to obtain the phase diagram and energy spectrum of the model. Next, in Sec.~\ref{dqpt} we investigate nonequilibrium dynamics of the model, using the notion of dynamical quantum phase transition. We argue that dynamical
quantum phase transitions occur in the quenches crossing a point, where the
%corresponding
propagating quasiparticles are massless. Finally,
the paper is summarized and concluded in Sec.~\ref{Sum}. Details of our results are presented in Appendices.

\section{The Model and Method}\label{Model}
The Hamiltonian of SSH model on 2D square lattice in the presence of staggered chemical potential is given as follow:
%%%%%%%%%%%%%%%%%%%%%%%%%%%%%%% Eq.1 %%%%%%%%%%%%%%%%%%%%%%%%%%%%%%%
\small
\begin{eqnarray}
\no
\mathcal{H}=\sum\limits_{n,m}[&&wc^{\dagger}_{2n-1,m}c_{2n,m}+vc^{\dagger}_{2n,m}c_{2n+1,m}+wc^{\dagger}_{n,2m-1}c_{n,2m}\\
\label{eq1}
&&+vc^{\dagger}_{n,2m}c_{n,2m+1}+h.c)+\mu(-1)^{n+m}c^{\dagger}_{n,m}c_{n,m}]
\end{eqnarray}
\normalsize
%%%%%%%%%%%%%%%%%%%%%%%%%%%%%%%%%%%%%%%%%%%%%%%%%%%%%%%%%%%%%%%%%
where $c^{\dagger}_{n,m}$, and $c_{n,m}$, are fermion creation and annihilation operators, $w$, $v$,
are alternatinvg hopping amplitudes and $\mu$ is chemical potential (see Appendix.~\ref{ap-A}).
The presence of two types of hopping in each direction can be regarded as splitting the square lattice into plaquettes formed by either $w$- or $v$-bonds.
Introducing Nambu spinor $\mathds{C}^{\dagger}=(c^{A\dagger}_{\textbf{k}}, c^{B\dagger}_{\textbf{k}}, c^{C\dagger}_{\textbf{k}}, c^{D\dagger}_{\textbf{k}})$,
where, $c^{\nu\dagger}_{\textbf{k}}$, $c^{\nu}_{\textbf{k}}$, ($\nu=A, B, C, D$) represent creation and annihilation fermion operator in each plaqutte (see Appendix.~\ref{ap-A}), the Fourier transformed Hamiltonian can be expressed as
%
%%%%%%%%%%%%%%%%%%%%%%%%%%%%%%% Eq.2 %%%%%%%%%%%%%%%%%%%%%%%%%%%%%%%
\begin{eqnarray}
\mathcal{H}=\sum_{\textbf{k}}{\mathds{C}^{\dagger}_{\textbf{k}}H(\textbf{k})\mathds{C}_{\textbf{k}}}.
\label{eq2}
\end{eqnarray}
%%%%%%%%%%%%%%%%%%%%%%%%%%%%%%%%%%%%%%%%%%%%%%%%%%%%%%%%%%%%%%%%%
%
The Bloch Hamiltonian $H(\textbf{k})$ in Eq. (\ref{eq2}), is $H(\textbf{k})=(w+v\cos(k_{x}))\Gamma_{01}+v\sin(k_{x})\Gamma_{32}+(w+v\cos(k_{y}))\Gamma_{11}-v\sin(k_{y})\Gamma_{21}+\mu\Gamma_{03}$,
with $\Gamma_{ij}=\sigma_{i}\otimes\sigma_{j}$, where $\sigma_{1,2,3}$, are Pauli matrices and $\sigma_{0}$, is $2\times2$
identity matrix.

By diagonalizing $H(\textbf{k})$ the quasiparticle Hamiltonian obtains as follow
$\mathcal{H}=\sum_{\alpha=1}^{4}\sum_{\textbf{k}}\varepsilon_{\alpha, \textbf{k}}\beta_{\alpha, \textbf{k}}^{\dag}\beta_{\alpha, \textbf{k}}$, with $\beta_{\alpha, \textbf{k}}^{\dag}$ and $\beta_{\alpha, \textbf{k}}$ are
quasi-fermions as linear combinations of the elements in the Nambu spinor with the corresponding energy bands
\begin{eqnarray}
\varepsilon_{1, \textbf{k}}=-\varepsilon_{4, \textbf{k}}=-\sqrt{(\sqrt{\gamma_{x}(k_{x})}+\sqrt{\gamma_{y}(k_{y})})^{2}+\mu^{2}}, \nonumber \\
\label{eq3}\varepsilon_{2, \textbf{k}}=-\varepsilon_{3, \textbf{k}}=-\sqrt{(\sqrt{\gamma_{x}(k_{x})}-\sqrt{\gamma_{y}(k_{y})})^{2}+\mu^{2}},
\end{eqnarray}
where $\gamma_{l}(k_l)=w^{2}+v^{2}+2wv\cos(k_l)$, ($l=x,y$). The ground state $|\Psi_0\rangle$
is obtained by filling up the negative-energy quasi-fermion states,
$|\Psi_0\rangle = \prod_\textbf{k} \beta_{2, \textbf{k}}^{\dag} \beta_{1, \textbf{k}}^{\dag} |V\rangle$, where $|V\rangle$ is the Bogoliubov vacuum annihilated by the $\beta_\textbf{k}$'s (see Appendix.~\ref{ap-A}).

\begin{figure}[htp]
\includegraphics[width=0.49\columnwidth]{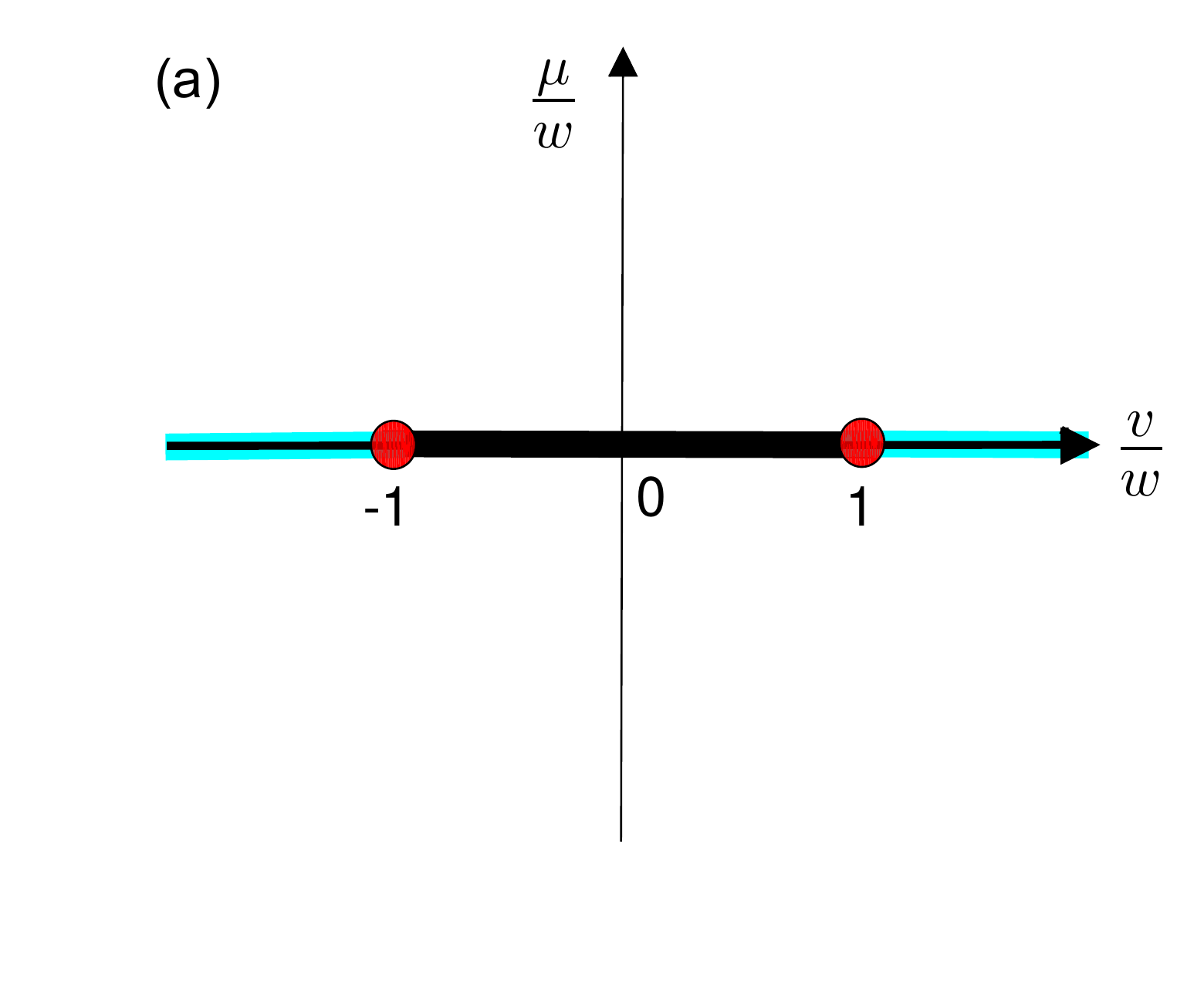}
\includegraphics[width=0.49\columnwidth]{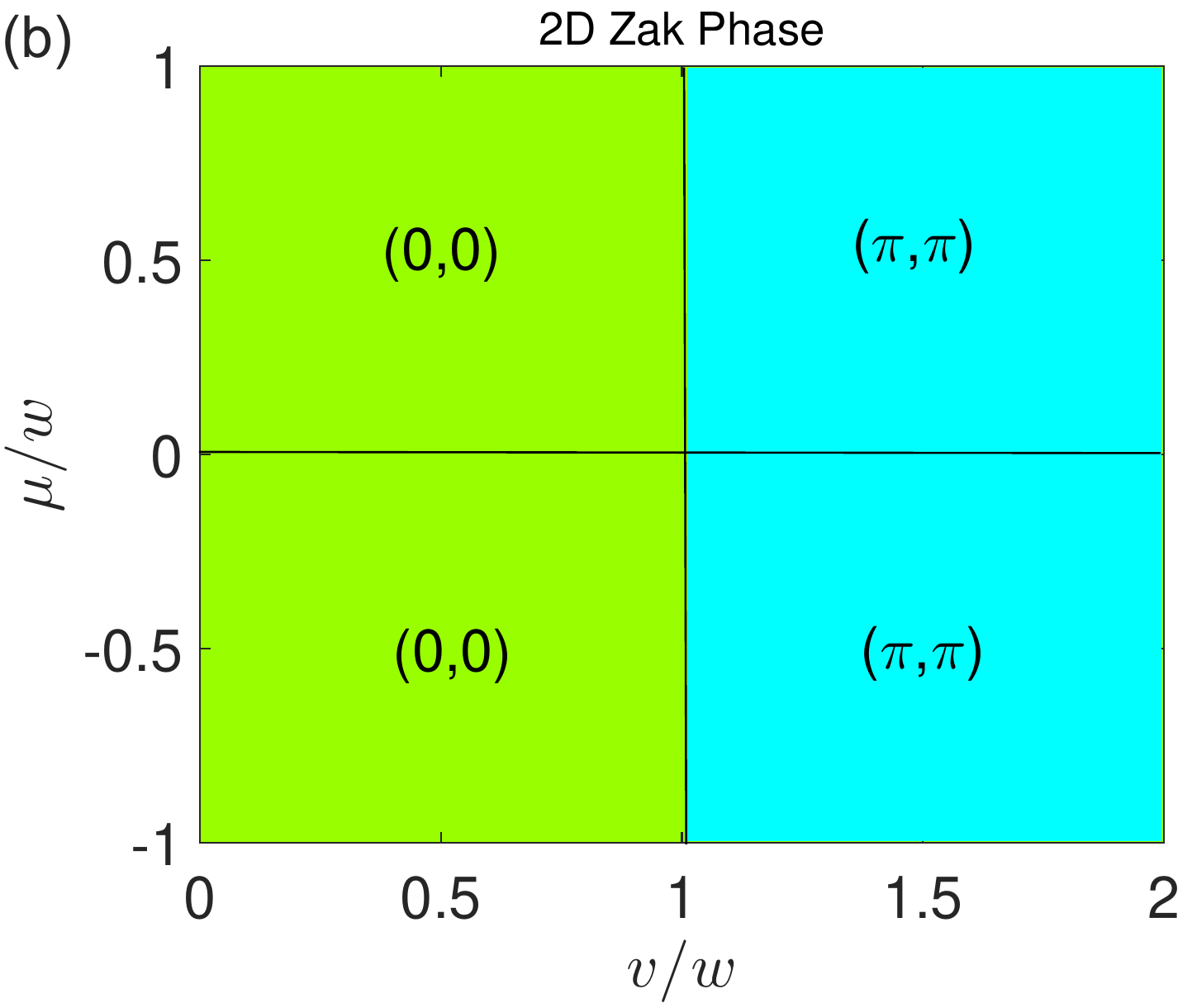}
\caption{(Color online) (a) Phase diagram of 2D SSH model in the presence of a staggered
chemical potential. The line $\mu=0$ is massless critical line with macroscopic degeneracy, where
DQPT happens only when quenches cross the red circles ($|v/w|=1$).
(b) Density plot of 2D Zak phase of extended SSH model versus $v/w$ and $\mu/w$.
The 2D Zak Phase is zero $\mathcal{Z}_x=\mathcal{Z}_y=0$ for $v<w$ in the absence/presence of
the staggered chemical potential $\mu$. Although, the 2D Zak Phase has a finite
value $\mathcal{Z}_x=\mathcal{Z}_y=\pi$ for $v>w$, in the absence/presence of the staggered
chemical potential ($\mu$) the phases of system is not topological for $\mu\neq 0$.}
\label{fig2}
\end{figure}

Accordingly, the quasiparticle spectrum of 2D SSH model on the square lattice is plotted in Fig. \ref{fig1}, in the presence/absence of staggered chemical potential and at/away from the isotropic point (IP) $w=v=1$. For $\mu\neq0$, the system is always gapped regardless the values of $v/w$, which is observed in Figs.\ref{fig1}(a)-(b). On the contrary, and as required for the existence of the quantum critical line $\mu=0$, the energy gap between $\varepsilon_{2,\textbf{k}}$ and $\varepsilon_{3,\textbf{k}}$ closes for $|k_{x}|=|k_{y}|$, at $\mu=0$, for any arbitrary ratio of $v/w$ as is shown in Figs.\ref{fig1}(c)-(d). Furthermore, at IP, the energy gap between $\varepsilon_{1,\textbf{k}}$ and $\varepsilon_{4,\textbf{k}}$ closes at $k=\pi$ (Fig. \ref{fig1}(d)) while it is nonzero away from IP (Fig. \ref{fig1}(c)). Hence, $\mu=0$ defines the line of EQPT between two distinct insulator phases.  Although all points on the line $\mu=0$ are critical ones,  a topological phase transition occurs only at the highly symmetric point $|v/w|=1$ (Fig. \ref{fig2}(a)), beyond which the doubly degenerate edge states appear \cite{Liu2017}. The nontrivial topological phase was characterized by the 2D Zak phase \cite{Zak1989} (see Appendix.~\ref{ap-B}) accompanying a fractional charge polarization in each direction \cite{Obana2019} (Fig. \ref{fig2}(b)). The robustness of edge states is determined by the difference between inter-plaquette and intra-plaquette hopping amplitudes ($|v-w|$). In the presence of staggered chemical potential, the edge states always exist in the band gaps as far as the amplitude of chemical potential is smaller than $|v-w|$, without any gap closing at $v=w$ (see Appendix.~\ref{ap-B}). These edge states are not robust against a small local
perturbation in contrast to the real boundary edge states at $\mu=0$, hence, the phase of system for $\mu\neq 0$
belongs to a non-topological phase (see Appendix.~\ref{ap-B}).
It is worthwhile to note that, on the critical line $\mu=0$, the structure of the Hilbert space gives a macroscopic degeneracy of $2^{2N-1}$ away from IP, and an enhanced degeneracy of $2^{2N+3}$ at IP.
%
%%%%%%%%%%%%%%%%%%%%%%%  Fig.2   %%%%%%%%%%%%%%%%%%%%%%%
%%%%%%%%%%%%%%%%%%%%%%%%%%%%%%%%%%%%%%%%%%%%%%%%%%%%%%%%

\begin{comment}
Furthermore, we investigate the same quench for non-zero chemical potential, where the system possesses nontrivial charge polarization
while the energy band gap does not close . We also investigate the DQPT by quenching the staggered chemical potential through the massless critical line.
\end{comment}

\section{DQPT and energy band gap}\label{dqpt}
In what follows, first we study quenches across TQPT point by sudden quench of
inter-plaquette hopping amplitude $v$, along the massless critical line $\mu=0$ (we set $w=1$ as the scale of energy
in our numerical simulations). To calculate the LO for 2D SSH model, we imagine that
the system is initially prepared in its ground state $|\Psi_0\rangle$.
A straightforward (though lengthy) calculation gives the complete set of eigenstates
of the model from which an exact expression for the LO can be extracted.
Quenching the inter-plaquette hopping amplitude from $v_{1}$ to $v_{2}$ at $\mu=0$, we obtain
%
%%%%%%%%%%%%%%%%%%%%%%%%%%%%%%%% Eq.4 %%%%%%%%%%%%%%%%%%%%%%%%%%%%%%%%%%%%%
%{\small
\bea
\label{eq4}
{\cal L} &&(v_1,v_2,t)\!=\!\langle\Psi_0(v_1)|\exp(-i\mathcal{H}(v_2)t)|\Psi_0(v_1)\rangle\\
\no
\!&&=\!\prod_{\textbf{k}}\!{\cal L}_\textbf{k}(v_1,v_2,t)=\prod_{\textbf{k}}\Big[(\frac{b+a}{2b})^{2}e^{-i\epsilon_{0,\textbf{k}}(v_{2})t}\\
\no
\!&&+ \!(\frac{b-a}{2b})^{2}e^{i\epsilon_{0,\textbf{k}}(v_{2})t}+\frac{c^2}{2b^2}\Big],
\eea
%}
%%%%%%%%%%%%%%%%%%%%%%%%%%%%%%%%%%%%%%%%%%%%%%%%%%%%%%%%%%%%%%%%%%%%%%
where, $a=1+v_1v_2+(v_1+v_2)\cos(k_{x})$, $b=\epsilon_{0,\textbf{k}}(v_{1})\epsilon_{0,\textbf{k}}(v_2)/4$, $c=(v_1-v_2)\sin(k_{x})$,
$\epsilon_{0, \textbf{k}}=\varepsilon_{1, \textbf{k}}+\varepsilon_{2, \textbf{k}}$ (see Appendix.~\ref{ap-C}).

\begin{figure*}[t!]
\centerline{
\includegraphics[width=0.25\linewidth]{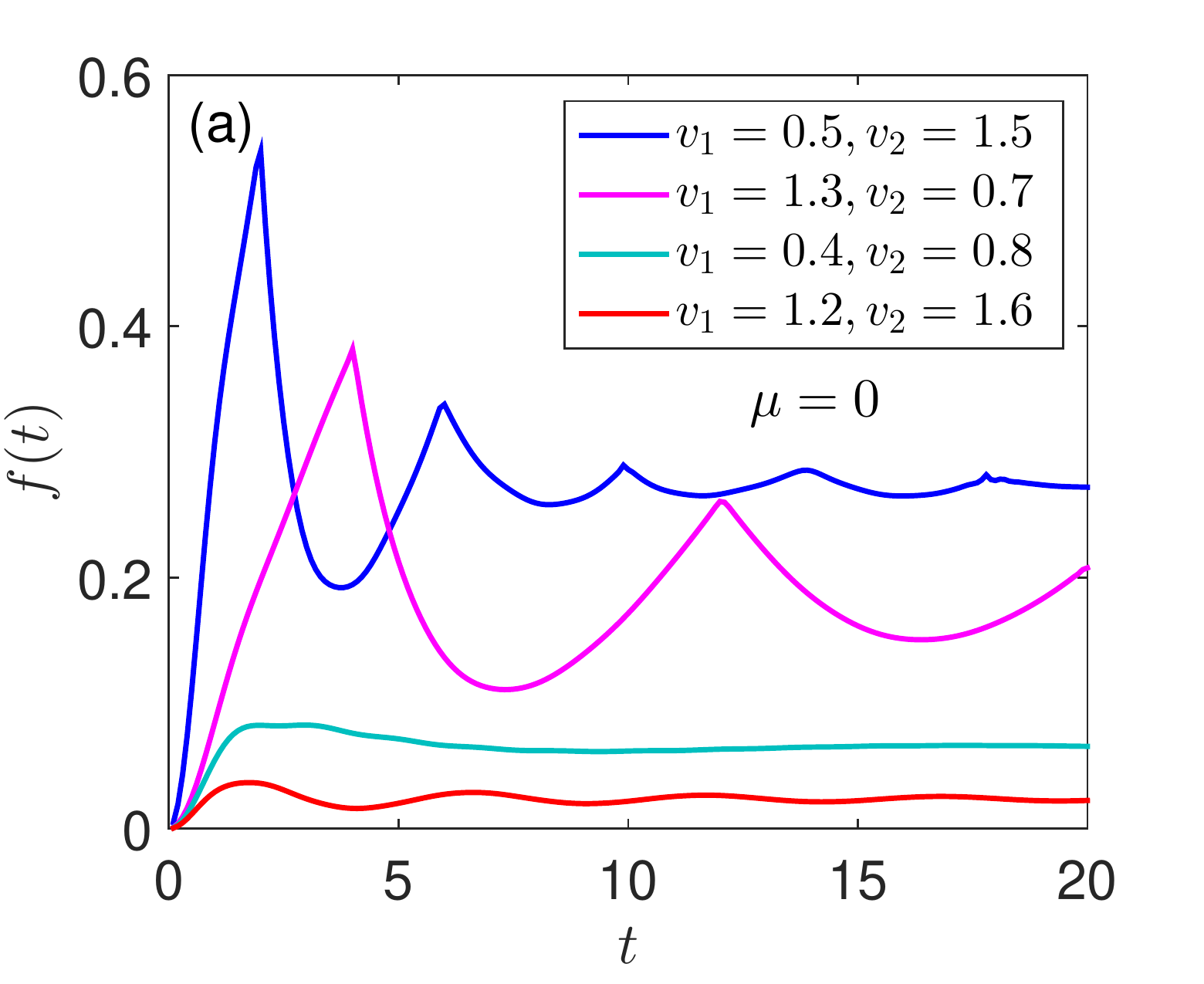}
\includegraphics[width=0.25\linewidth]{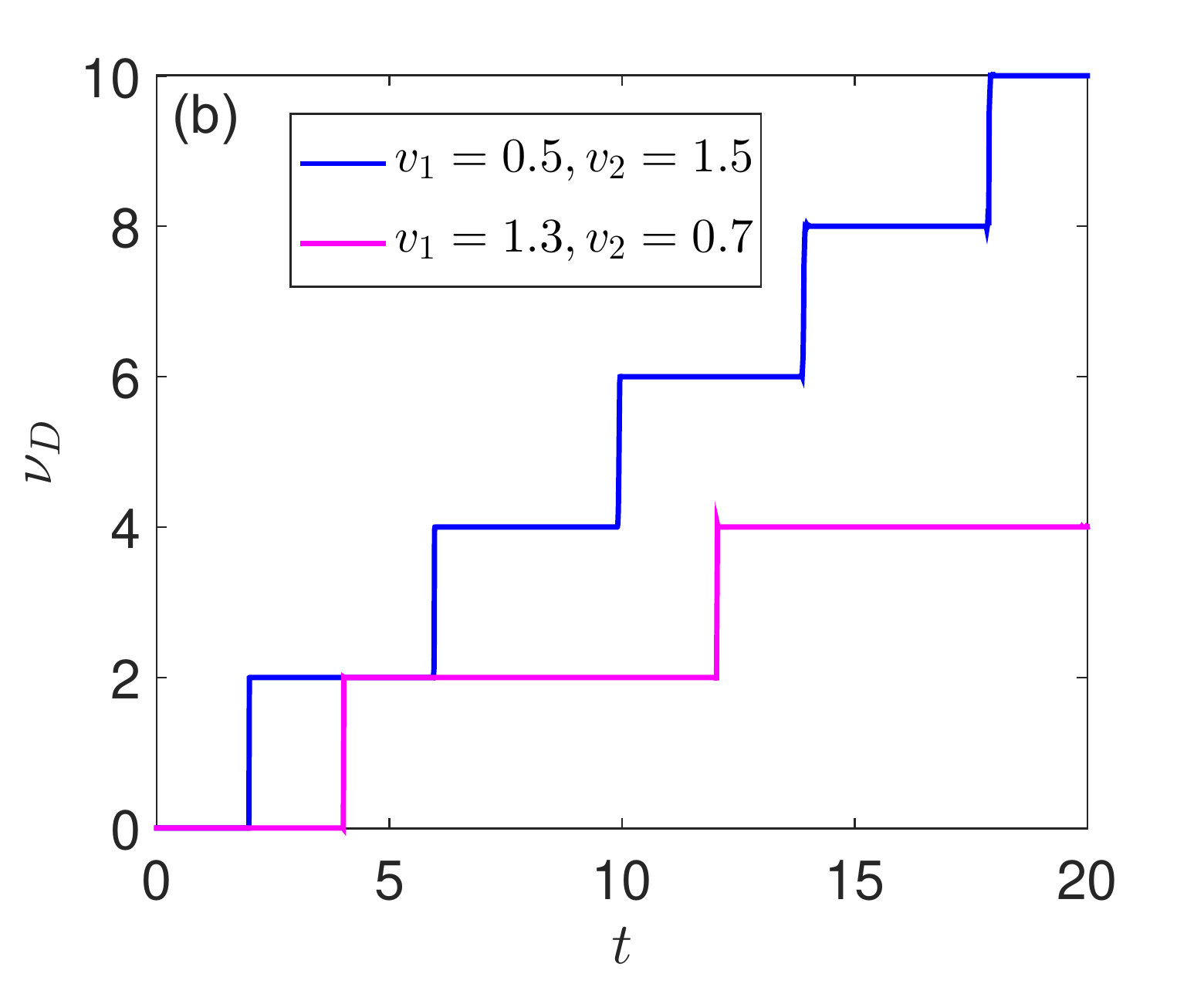}
\includegraphics[width=0.25\linewidth]{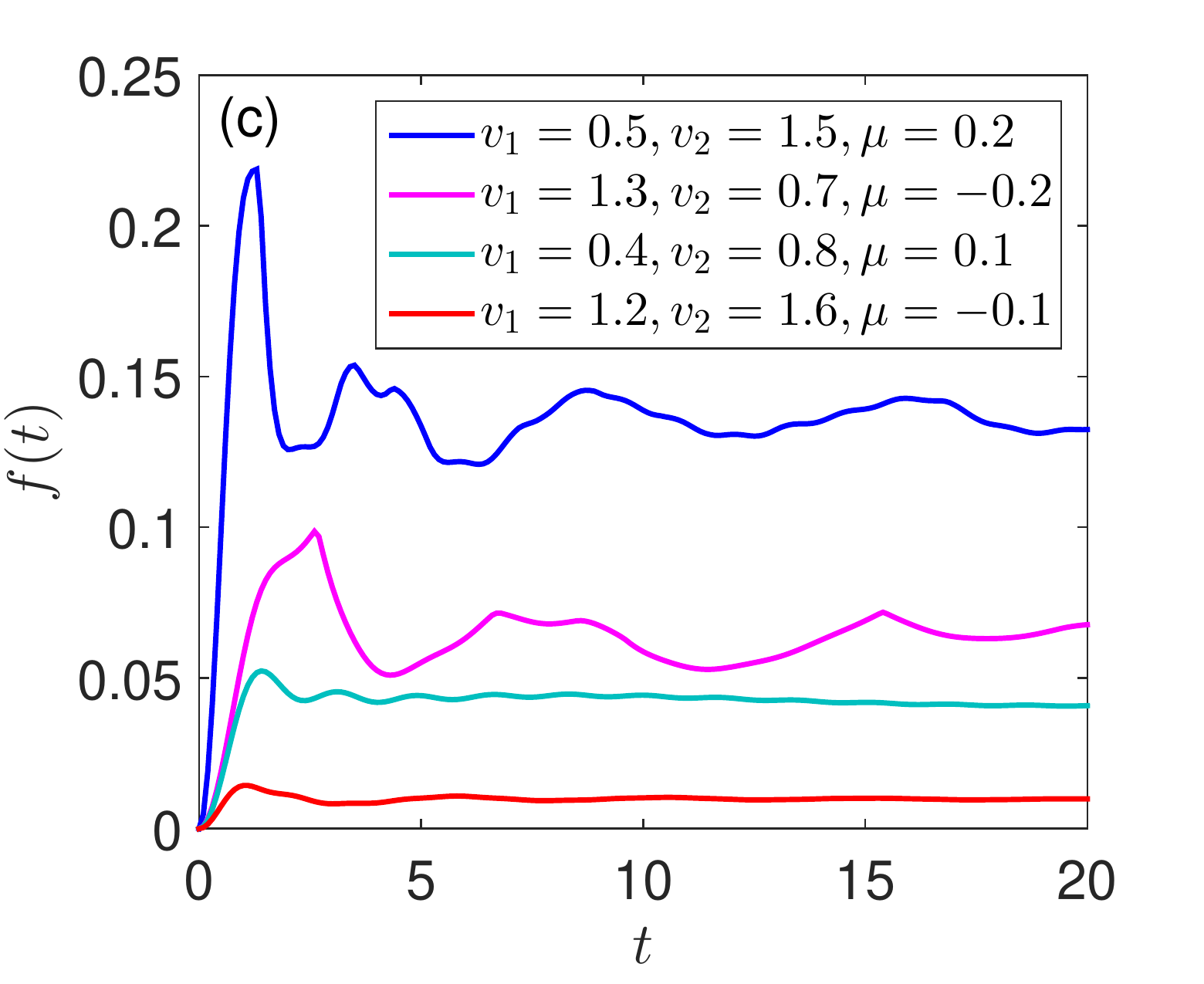}
\includegraphics[width=0.25\linewidth]{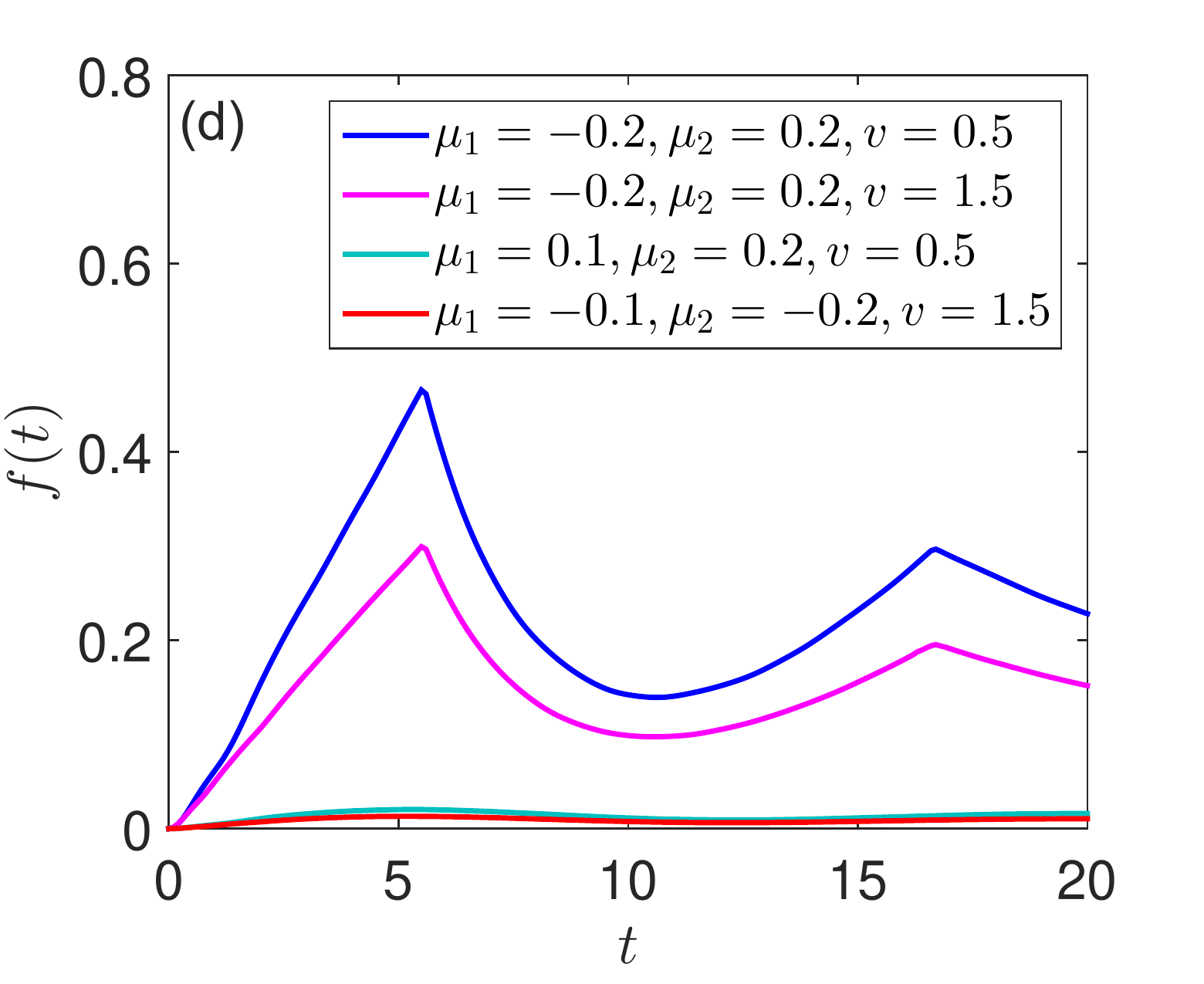}}
\caption{(Color online) (a) The free energy density $f(t)$, for different quenches of
the inter-plaquette hopping amplitude $v_1\rightarrow v_2$ at massless critical line $\mu=0$.
The Loschmidt overlap rate function shows periodic nonanalytic behaviors
for quenches across the topological quantum critical point $v=1$.
(b) The dynamical topological order parameter for quenches correspond to
Fig. \ref{fig3}(a). (c) The rate function of the LO in the presence of
chemical potential for quenches correspond to \ref{fig3}(a).
(d) The rate function of the LO for the sudden quenches of
the staggered chemical potential $\mu_1\rightarrow \mu_2$.
In all plots we set $w=1$.}
\label{fig3}
\end{figure*}

The rate function of LO, $f(t)=\frac{-1}{(2\pi)^2}Re[\int_{B.Z} d\textbf{k}\hspace{1mm}ln\mathcal{L}_\textbf{k}(t)]$ \cite{Mendl2019},
the dynamical counterpart of the free-energy density, has been plotted in Fig. \ref{fig3}(a) for quenches
at several inter-plaquette hopping ampliudes. The rate function
displays distinct (singular) behavior only for quenches across TQPT point $v=1$. Whenever the quench drives the system through the TQPT point, nonanalyticities
appear in the rate function of the LO.
For quenches within the nontrivial topological ($v_{1}, v_{2}>1$) or trivial ($v_{1}, v_{2}<1$)  phases, the dynamical free-energy density $f(t)$, shows completely analytic, smooth behavior.
Our analysis of Eq. (\ref{eq4}) manifests that, dynamical topological quantum phase transitions (DTQPTs) occur for any quench across the TQPT point at periodic critical times described by
%
%%%%%%%%%%%%%%%%%%%%%%%%%%%%%%%% Eq. %%%%%%%%%%%%%%%%%%%%%%%%%%%%%%%%%%%%%
\begin{eqnarray}
\no
t^{\ast}_n=(2n+1) t^{\ast},
\hspace{5mm} t^{\ast}=\frac{-\pi}{\epsilon_{0,\textbf{k}}(v_{2})}, \hspace{5mm} n\in \mathds{Z}.
\end{eqnarray}

%%%%%%%%%%%%%%%%%%%%%%%%%%%%%%%%%%%%%%%%%%%%%%%%%%%%%%%%%%%%%%%%%%%%%%
%

Emerging topological structure in the subsequent evolution of a quenched quantum system,
is characterised by a topological invariant $\nu_{D}(t)$ (see Appendix.~\ref{ap-D}), namely,
the dynamical topolocical order parameter (DTOP), which
is plotted in Fig. \ref{fig3}(b) corresponding to the behavior of Fig. \ref{fig3}(a). It is seen that,
DTOP gets quantized integer values, which changes by even values at every DTQPT
($t^{\ast}_n$) and uniquely indicates the dynamical phases between two DQPTs.
It should be mentioned that the change of two units in DTOP is originated from the ground state structure
of the model, which consists of two negative negative energy bands ($\varepsilon_{1, \textbf{k}}, \varepsilon_{2, \textbf{k}}$). Each band contributes to a unit change, which sums to two units for both bands.

%
%%%%%%%%%%%%%%%%%%%%%%%%%%%%%%%% Fig.3 %%%%%%%%%%%%%%%%%%%%%%%%%%%%%%%%%%%%%
%%%%%%%%%%%%%%%%%%%%%%%%%%%%%%%%%%%%%%%%%%%%%%%%%%%%%%%%%%%%%%%%%%%%%%
%

Secondly, we consider the dynamical free energy density, in the presence of staggered chemical potential ($\mu\neq0$), which has been depicted
in Fig. \ref{fig3}(c) for quenches correspond to Fig. \ref{fig3}(a). In this case, the rate function of the LO does not exhibit
signatures of DQPT, even for a quench crossing th IP at $\mu<|v_{1,2}-1|$, where the edge states always exist in the band gaps (see Appendix.~\ref{ap-B}) and the 2D Zak phase has a finite value for $v>1$.
Altough DQPT by a quench of inter-plaquette hopping amplitude is absent in the presence of staggered chemical potential, the dynamical free energy density shows nonanalyticities after a rapid quench of the staggered chemical potential across the highly degenerate critical line $\mu=0$, as shown in
Fig.\ref{fig3}(d).

Here, we are going to address under which circumstances DQPT takes place, which lies in the structure of LO.
Let us take a detailed look at the modulus of LO, which is called Loschmidt echo (LE), $L=|{\cal L}|^{2}$.
%%%%%%%%%%%%%%%%%%%%%%%%%%%%%%%%%%%%%%%%%%%%%%%%%%%%%%%%%%%%%%%%%%%%%%
In the following, we study LE for different types of quenches in the absence/presence of the chemical potential, respectively.

For a quench of inter-plaquette hopping amplitude ($v_1\rightarrow v_2$) at $\mu=0$, according to LO in Eq. (\ref{eq4}), LE is given by the following expression,
%%%%%%%%%%%%%%%%%%%%%%%%%%%%%%%%Eq.5%%%%%%%%%%%%%%%%%%%%%%%%%%%%%%%%%%%%%
\bea
\no
&L&=|{\cal L}(v_1,v_2,t)|^{2}=\prod_{\textbf{k}}L_{\textbf{k}}(v_1,v_2,t),\\
\no &L&_{\textbf{k}}(v_1,v_2,t)=1-A_{\textbf{k}}\sin^2(\varepsilon_{1, \textbf{k}}+\varepsilon_{2, \textbf{k}})t)\\
\label{eq5}
&&\;\;\;\;\;\;\;\;\;\;\;\;\;\;\;\;\;\;-B_{\textbf{k}}\sin^2(\varepsilon_{1, \textbf{k}}+\varepsilon_{2, \textbf{k}})t/2),
\eea
%%%%%%%%%%%%%%%%%%%%%%%%%%%%%%%%%%%%%%%%%%%%%%%%%%%%%%%%%%%%%%%%%%%%%%
%
where, $A_{\textbf{k}}$ and $B_{\textbf{k}}$ are products of linear combinations of the overlaps between $\textbf{k}$ modes of the
initial (ground) state of the pre-quenched Hamiltonian $\mathcal{H}(v_{1})$, and eigenstates of the post-quenched one $\mathcal{H}(v_{2})$ (see Appendix.~\ref{ap-C}).
It has to be mentioned that the real time instances at which DQPT appears is exactly equivalent to
the time instances at which at least one factor
in LO (LE) becomes zero i.e., $L_{\textbf{k}^{\ast}}=0$. It requires that the oscillating terms at $\textbf{k}^{\ast}$ reaches its maximum value equal to one.
An analysis reveals that the oscillation amplitudes $A_{\textbf{k}}$ and $B_{\textbf{k}}$ are small, except for $B_{\textbf{k}}$ in the case of quenching across the TQPT point, $v=1$.
This leads to having a zero in Eq. (\ref{eq5}) and consequently in $L$.
Hence, we expect the nonanalyticities occur at time instances at which $\sin^2(\epsilon_{0, \textbf{k}}(v_{2})t/2)=1$,
which results in $t^{\ast}_n=-(2n+1)\pi/\epsilon_{0, \textbf{k}}(v_{2})$.

However, in the presence of nonzero chemical potential ($\mu\neq0$), LE is generally given by,
%
%%%%%%%%%%%%%%%%%%%%%%%%%%%%%%%%% Eq.6 %%%%%%%%%%%%%%%%%%%%%%%%%%%%%%%%%%%%
\begin{eqnarray}
\no
&&L(v_1,v_2,\mu_{1},\mu_{2},t)=\\
\no
&&\prod_{\mathbf{k}}|1-A_{0,\mathbf{k}}\sin^{2}[(\varepsilon_{1,\mathbf{k}}(v_{2},\mu_{2})+\varepsilon_{2,\mathbf{k}}(v_{2}),\mu_{2})t]\\
\no
&&-A_{1,\mathbf{k}}\sin^{2}[(\varepsilon_{1,\mathbf{k}}(v_{2},\mu_{2})+\varepsilon_{2,\mathbf{k}}(v_{2},\mu_{2}))t/2]\\
\label{eq6}
&&-A_{2,\mathbf{k}}\sin^{2}[(\varepsilon_{1,\mathbf{k}}(v_{2},\mu_{2})-\varepsilon_{2,\mathbf{k}}(v_{2},\mu_{2}))t]\\
\no
&&-A_{3,\mathbf{k}}\sin^{2}[(\varepsilon_{1,\mathbf{k}}(v_{2},\mu_{2})-\varepsilon_{2,\mathbf{k}}(v_{2},\mu_{2}))t/2]\\
\no
&&-A_{4,\mathbf{k}}\sin^{2}[\varepsilon_{1,\mathbf{k}}(v_{2},\mu_{2})t]-A_{5,\mathbf{k}}\sin^{2}[\varepsilon_{2,\mathbf{k}}(v_{2},\mu_{2})t]|,
\end{eqnarray}
%%%%%%%%%%%%%%%%%%%%%%%%%%%%%%%%%%%%%%%%%%%%%%%%%%%%%%%%%%%%%%%%%%%%%%
%
%
%%%%%%%%%%%%%%%%%%%%%%%%%%%%%%%%% Eq.6 %%%%%%%%%%%%%%%%%%%%%%%%%%%%%%%%%%%%
%\begin{eqnarray}
%\no
%L(v_1,v_2,&&\mu_{1},\mu_{2},t)=\prod_{\textbf{k}}\Big[1-A_{0,\textbf{k}}\sin^{2}(\epsilon_{0, \textbf{k}}(v_{2},\mu_2)t)\\
%\no
%&&-A_{1,\textbf{k}}\sin^{2}(\epsilon_{0, \textbf{k}}(v_{2},\mu_2)t/2)\\
%\label{eq6}
%&&-A_{2,\textbf{k}}\sin^{2}(\epsilon_{1, \textbf{k}}(v_{2},\mu_2)t)\\
%\no
%&&-A_{3,\textbf{k}}\sin^{2}(\epsilon_{1, \textbf{k}}(v_{2},\mu_2)t/2)\\
%\no
%&&-A_{4,\textbf{k}}\sin^{2}[(\epsilon_{0, \textbf{k}}(v_{2},\mu_2)+\epsilon_{1, \textbf{k}}(v_{2},\mu_2))t/2]\\
%\no
%&&-A_{5,\textbf{k}}\sin^{2}[(\epsilon_{0, \textbf{k}}(v_{2},\mu_2)-\epsilon_{1, \textbf{k}}(v_{2},\mu_2))t/2]\Big],
%\end{eqnarray}
%%%%%%%%%%%%%%%%%%%%%%%%%%%%%%%%%%%%%%%%%%%%%%%%%%%%%%%%%%%%%%%%%%%%%%
%
where,
%$\epsilon_{0,\textbf{k}}=\varepsilon_{1,\textbf{k}}+\varepsilon_{2,\textbf{k}}$, $\epsilon_{1,\textbf{k}}=\varepsilon_{1,\textbf{k}}-\varepsilon_{2,\textbf{k}}$, and
$A_{j,\textbf{k}},~j=0,\cdots5$ are products of linear combinations of the overlaps between $\textbf{k}$ modes of the initial and post-quenched eigenstates (see Appendix.~\ref{ap-C}).
%(ground) state of the pre-quenched Hamiltonian $\mathcal{H}(\mu,w,v_{1})$, and eigenstates of the post-quenched one $\mathcal{H}(\mu,w,v_{2})$ (see Supplemental Material \cite{Sadrzadeh}).
We find that all oscillation amplitudes are small for a quench of inter-plaquette hopping amplitude from $v_1\rightarrow v_2$ at $\mu_1=\mu_2$
across the isotropic point, $v=1$. In this case, no mode contributes
destructively in LE and consequently rate function of LO shows completely analytic, smooth behavior (Fig.\ref{fig3}(c)).

On the other hand, if we consider a quench of the staggered chemical potential from $\mu_{1}\rightarrow \mu_{2}$ at $v_1=v_2$,
whenever the model is quenced through the massless critical line $\mu=0$, away from the isotropic point, all oscillation amplitudes
are small except $A_{5,\textbf{k}}$, which can reaches its maximum value.
It leads to get a mode, which contributes destructively to LE to have nonanalyticities in the rate function of LO (Fig.\ref{fig3}(d)).

To understand the origin of distinct behaviors of LE for different quenches mentioned above, recall
that the nonanalyticities are controlled by quasiparticles in the lowest energy bands, $\varepsilon_{1, \textbf{k}}$, and $\varepsilon_{2, \textbf{k}}$. Oscillation amplitudes are function of the occupation probability of particle-hole excitations, which are created
in a sudden quench. In the presence of the staggered chemical potential ($\mu\neq0$), the system is always gapped, separating $\varepsilon_{1, \textbf{k}}$ and $\varepsilon_{2, \textbf{k}}$ from the higher bands (Fig. \ref{fig1}(a)-(b)). Thus, in a quench from $v_1\rightarrow v_2$ at $\mu\neq0$, quasiparticle excitations are hindered and, as a consequence, it diminishes the oscillation amplitudes in LE (Eq. (\ref{eq6})).

However, the second filled band in the ground state, $\varepsilon_{2, \textbf{k}}$ collapses to zero and becomes dispersionless for any arbitrary ratio of $v/w$ at $\mu=0$ for $|k_{x}|=|k_{y}|$ (Fig. \ref{fig1}(c)). It is therefore expected that, for any quench of inter-plaquette hopping amplitude $v_1\rightarrow v_2$ at $\mu=0$, dispersionless quasiparticles of $\varepsilon_{2, \textbf{k}}$ band at $\mu=0$, contribute significantly to the time-dependent parts of the oscillation terms in Eq. (\ref{eq5}). But the nonanalyticities of the rate function of LO at $\mu=0$ does not happen until quenching across the IP $v=1$ (Fig. \ref{fig3}(a)), where the gap of $\varepsilon_{1, \textbf{k}}$ band closes at $k=\pi$ (Fig. \ref{fig1}(d)). Hence, there are dispersionless quasiparticles of $\varepsilon_{1, \textbf{k}}$ band which yield $B_{\textbf{k}}$ becomes maximum in the case of quenching across the TQPT point, (presented in Eq. (\ref{eq5})). However, in the absence of the staggered chemical potential and away from the isotropic point, $\varepsilon_{1, \textbf{k}}$ remains gapped for all $\textbf{k}$ (Fig. \ref{fig1}(c)), hence holding back quasiparticle excitations from that band. Therefore, quenching the staggered chemichal potential away from the IP and across the critical line $\mu=0$, yields a large oscillation amplitude due to dispersionless quasiparticles of $\varepsilon_{2, \textbf{k}}$ band for $|k_{x}|=|k_{y}|$ at $\mu=0$. This explains why $A_{5,\textbf{k}}$ becomes maximum in the case of quenching across the critical line $\mu=0$, in Eq. (\ref{eq6}), which results in nonanalyticities of the rate function of LO (Fig. \ref{fig3}(d)).\\

\section{Summary and conclusion}\label{Sum}
We have studied the quantum quench of the two dimensional SSH model in
the presence of the staggered chemical potential to address
the connection between dynamical quantum phase transition
and the absence of an energy gap in the elementary excitations of the Hamiltonian.
By examining how the eigenstates of the models imprint the Loschmidt echo, we find that dynamical
quantum phase transitions occur in the quenches crossing a point, where the corresponding quasiparticles,
having an impact on the Loschmidt overlaps, are massless. Although a quantum phase transition generically
supports massless excitations, our case study reveals
that these excitations may not necessarily couple to  the dynamical quantum phase transition.
In other words, dynamical phase transition is absent if the quasiparticles, which control the Loshmidt overlap
remain fully gapped even if the quantum critical points crosses during the quench.
We have also shown that, the 2D Zak Phase has a finite value in the presence of a staggered chemical potential while
there is no TQPT, which reveals that 2D Zak phase is not the proper measure
to describe the TQPT in 2D SSH model.
However, dynamical topolocical order parameter can truly capture the TQPT on the zero Berry curvature line, where the Chern number is zero.
We would like to suggest that our findings can be explored experimentally using a two dimensional
lattice of nanoelectromechanical resonators \cite{PhysRevB.100.024310,PhysRevB.101.174303}(see Appendix \ref{ap-E}).
Moreover, these results can be examined using the multibands models, especifically the
multibands metalic systems.
\appendix
\section{Eigenstates and eigenvalues of the SSH model \label{ap-A}}
\begin{figure}
\includegraphics[width=0.7\columnwidth,height=0.7\columnwidth]{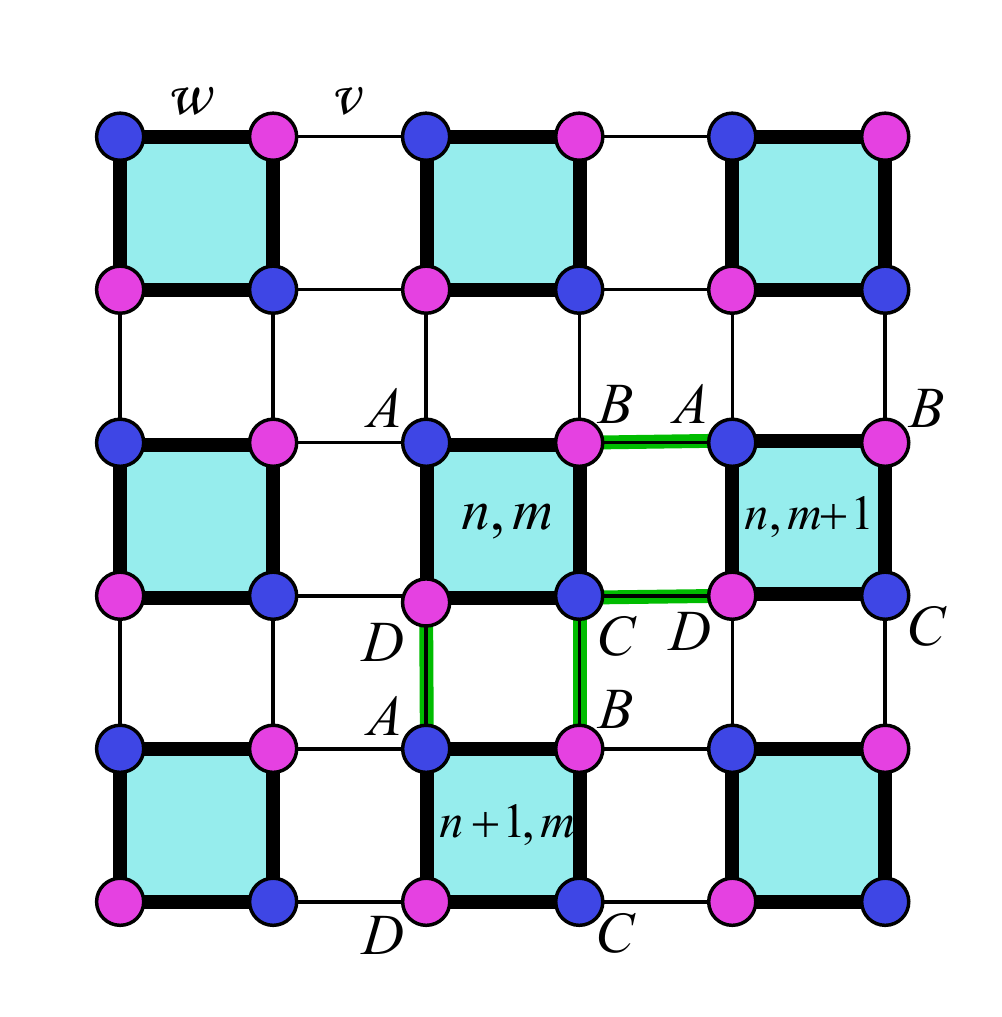}
\caption{ 2D SSH model on a square lattice. Thick and thin black bonds are
$w$ and $v$ hopping amplitudes, respectively. There is a staggered chemical
potential $\mu$ on each lattice site.}
\label{figS1}
\end{figure}

The presence of two types of hoppings in each direction in the two dimensional (2D) SSH Hamiltonian $\mathcal{H}$ in Eq. (\ref{eq1}) can be regarded as splitting the square lattice into plaquettes. Thus, there are four atoms in each plaquette labeled by $A, B, C, D$ (Fig. \ref{figS1}). In this case,
the Hamiltonian can be written in the following form
%%
%%%%%%%%%%%%%%%%%%%%%%%%  Fig.S0   %%%%%%%%%%%%%%%%%%%%%%%
%%%%%%%%%%%%%%%%%%%%%%%%%%%%%%%%%%%%%%%%%%%%%%%%%%%%%%%%%%%%%%
%
%
%%%%%%%%%%%%%%%%%%%%%%%%%%%%%%% Eq. S1 %%%%%%%%%%%%%%%%%%%%%%%%%%%%%%%%
\begin{eqnarray}
\no
\mathcal{H} &=& \displaystyle\sum\limits_{n} \Big[w(c^{A\dagger}_{n,m}c^B_{n,m}+c^{B\dagger}_{n,m}c^C_{n,m}+c^{C\dagger}_{n,m}c^D_{n,m}+c^{D\dagger}_{n,m}c^A_{n,m})\\
\no &+&v(c^{B\dagger}_{n,m}c^A_{n,m+1}+c^{C\dagger}_{n,m}c^D_{n,m+1}+c^{D\dagger}_{n,m}c^A_{n+1,m}+c^{C\dagger}_{n,m}c^B_{n+1,m})\\
\nonumber
&+& \frac{\mu}{2}(c^{A\dagger}_{n,m}c^A_{n,m}-c^{B\dagger}_{n,m}c^B_{n,m}+c^{C\dagger}_{n,m}c^C_{n,m}-c^{D\dagger}_{n,m}c^D_{n,m})+h.c\Big].\\
\label{eqS1}
\end{eqnarray}
%%%%%%%%%%%%%%%%%%%%%%%%%%%%%%%%%%%%%%%%%%%%%%%%%%%%%%%%%%%%%%%%%%%%%%%%
%

By Fourier transforming the SSH Hamiltonian $\mathcal{H}$, Eq. (\ref{eqS1}), is transformed into a sum of commuting Hamiltonian terms, $H_{\textbf{k}}$, each describing a different $k$ mode,

%
%%%%%%%%%%%%%%%%%%%%%%%%%%%%%%% Eq. S2 %%%%%%%%%%%%%%%%%%%%%%%%%%%%%%%%
\begin{eqnarray}
\no
H_{\textbf{k}}&&=\Big{[(w+ve^{-ik_x})c^{A\dagger}_{\textbf{k}}c^B_{\textbf{k}}+(w+ve^{ik_y})c^{B\dagger}_{\textbf{k}}c^C_{\textbf{k}}}\\
\no &&+(w+ve^{ik_x})c^{C\dagger}_{\textbf{k}}c^D_{\textbf{k}}+(w+ve^{-ik_y})c^{D\dagger}_{\textbf{k}}c^A_{\textbf{k}}\\
\no &&+(w+ve^{ik_x})c^{B\dagger}_{\textbf{k}}c^A_{\textbf{k}}+(w+ve^{-ik_y})c^{C\dagger}_{\textbf{k}}c^B_{\textbf{k}}\\
\no &&+(w+ve^{-ik_x})c^{D\dagger}_{\textbf{k}}c^C_{\textbf{k}}+(w+ve^{ik_y})c^{A\dagger}_{\textbf{k}}c^D_{\textbf{k}}\\
&&+\mu(c^{A\dagger}_{\textbf{k}}c^A_{\textbf{k}}-c^{B\dagger}_{\textbf{k}}c^B_{\textbf{k}}+c^{C\dagger}_{\textbf{k}}c^C_{\textbf{k}}-c^{D\dagger}_{\textbf{k}}c^D_{\textbf{k}})\Big],
\label{eqS2}
\end{eqnarray}
%%%%%%%%%%%%%%%%%%%%%%%%%%%%%%%%%%%%%%%%%%%%%%%%%%%%%%%%%%%%%%%%%%%%%%%%
%
where, $c^{\nu\dagger}_{\textbf{k}}$, $c^{\nu}_{\textbf{k}}$, ($\nu=A, B, C, D$) represent creation and annihilation fermion operator in each plaqutte. Eigenstates and eigenvalues of 2D SSH model, in the presence of the staggered chemical potential, is obtained by diagonalizing each mode $H_{\textbf{k}}$ of Eq. (\ref{eqS2}), independently. This is done in two ways: First using a generalized Bogoliubov transformation, which maps $H_{\textbf{k}}$ onto the Bogoliubov quasiparticle Hamiltonian $H(\textbf{k})$, $H_{\textbf{k}}=\mathds{C}^{\dagger}_{\textbf{k}}H(\textbf{k})\mathds{C}_{\textbf{k}}$, where, $\mathds{C}^{\dagger}_{\textbf{k}} = (c^{A\dagger}_{\textbf{k}}, c^{B\dagger}_{\textbf{k}}, c^{C\dagger}_{\textbf{k}},c^{D\dagger}_{\textbf{k}})$, and

\begin{eqnarray}
\nonumber
\label{eqS3}H(\textbf{k}) =
\begin{pmatrix}
 \mu & w+ve^{-ik_x}& 0& w+ve^{ik_y}\\
 w+ve^{ik_x}& -\mu & w+ve^{ik_y}& 0\\
 0& w+ve^{-ik_y}& \mu & w+ve^{ik_x}\\
 w+ve^{-ik_y}& 0& w+ve^{-ik_x}& -\mu
\end{pmatrix}.
\nonumber\\
\end{eqnarray}
The diagonalization of $H(\textbf{k})$ leads to the quasiparticle representation of Hamiltonian obtained as follow
$\mathcal{H}=\sum_{\alpha=1}^{4}\sum_{\textbf{k}}\varepsilon_{\alpha, \textbf{k}}\beta_{\alpha, \textbf{k}}^{\dag}\beta_{\alpha, \textbf{k}}$. Here, the quasiparticle operators $\beta_{\textbf{k}}^{\alpha}$ and $\beta_{\textbf{k}}^{\alpha \dagger}, \alpha=A, B, C,D$, expressed in terms of the fermion operators of Eq. (\ref{eqS2}) have the corresponding energy bands $\varepsilon_{\alpha, \textbf{k}}$.
Assuming one electron per lattice site, the many-particle groundstate of 2D SSH Hamiltonian is obtained by filling up the two lowest bands, $\varepsilon_{1,\textbf{k}}$ and $\varepsilon_{2,\textbf{k}}$.

%%%%%%%%%%%%%%%%%%%%%%%  Fig.S3   %%%%%%%%%%%%%%%%%%%%%%%
\begin{figure*}
\centerline{
\includegraphics[width=0.33\linewidth]{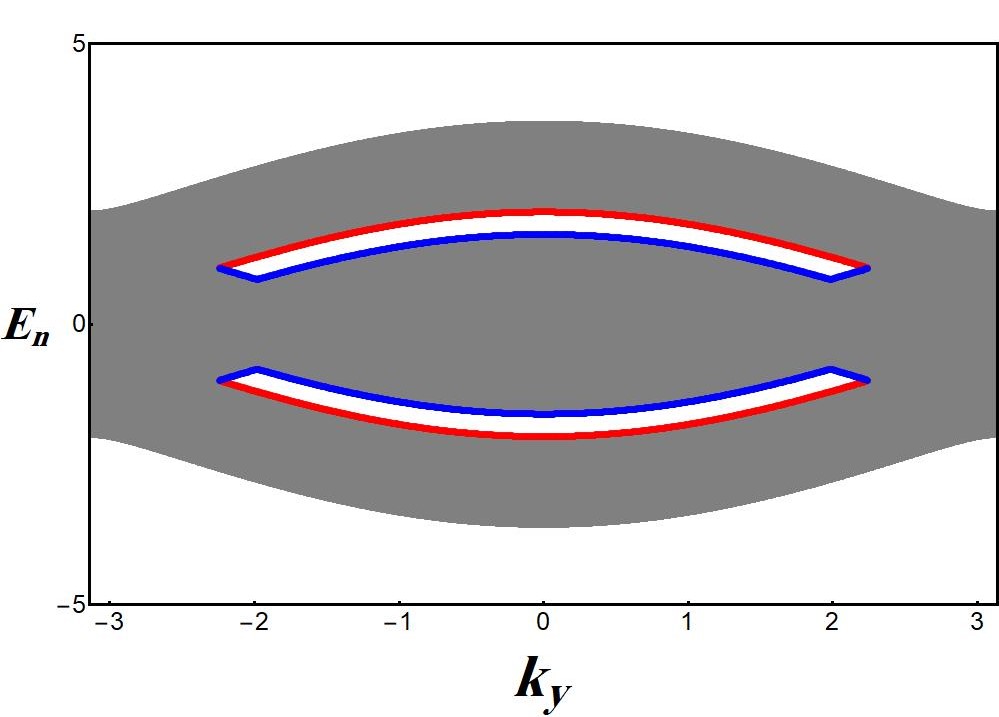}
\includegraphics[width=0.33\linewidth]{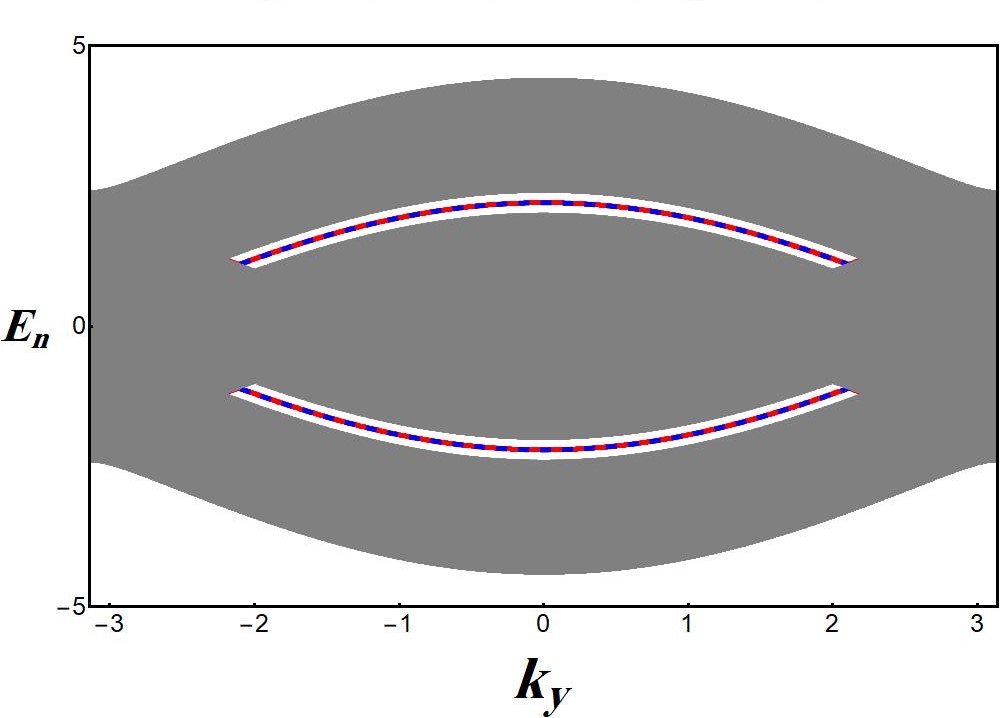}
\includegraphics[width=0.33\linewidth]{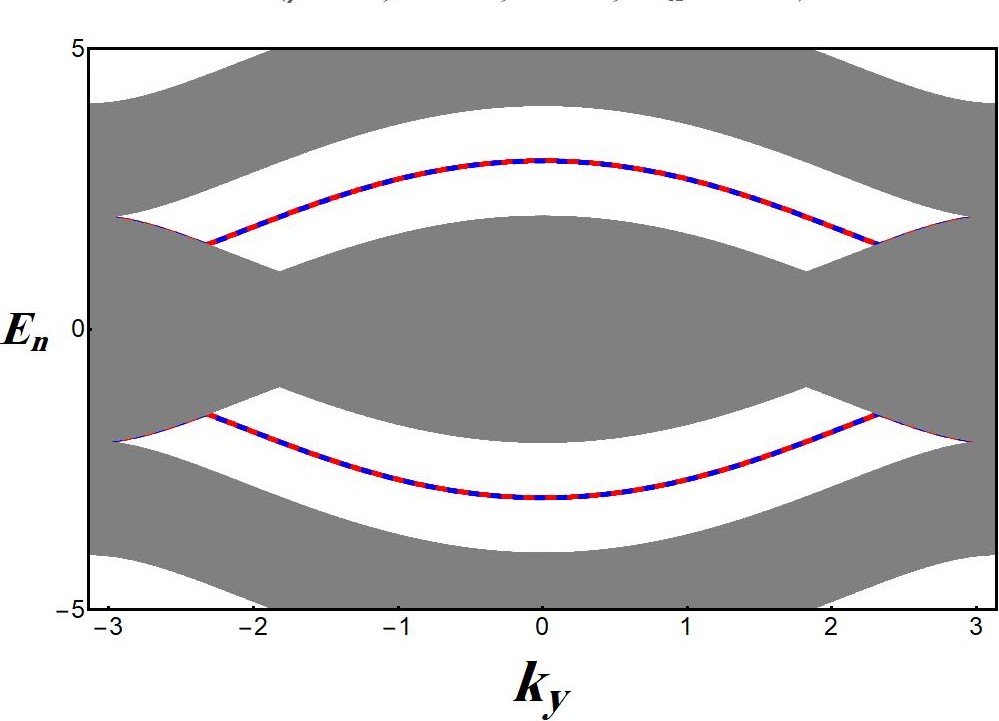}}
\caption{(Color online) Spectrum of SSH ribbon along x direction for $\mu=0$, $w=1$ and $N_x=225$,
(a) $v=0.8$, (b) $v=1.2$, and $v=2$. As seen double degenerate edge states naturally appear when
the intra-plaquette hopping is smaller than inter-plaquette hopping $v>w$.}
\label{figS2}
\end{figure*}
%%%%%%%%%%%%%%%%%%%%%%%%%%%%%%%%%%%%%%%%%%%%%%%%%%%%%%%
%
%
%%%%%%%%%%%%%%%%%%%%%%%  Fig.S5   %%%%%%%%%%%%%%%%%%%%%%%
\begin{figure*}
\centerline{
\includegraphics[width=0.33\linewidth]{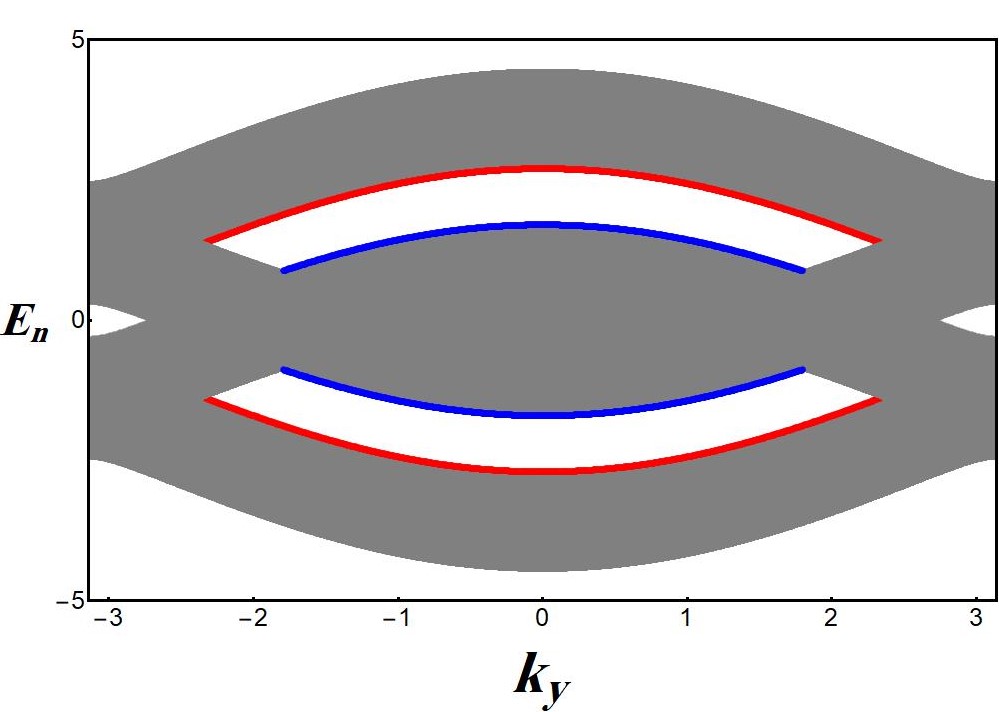}
\includegraphics[width=0.33\linewidth]{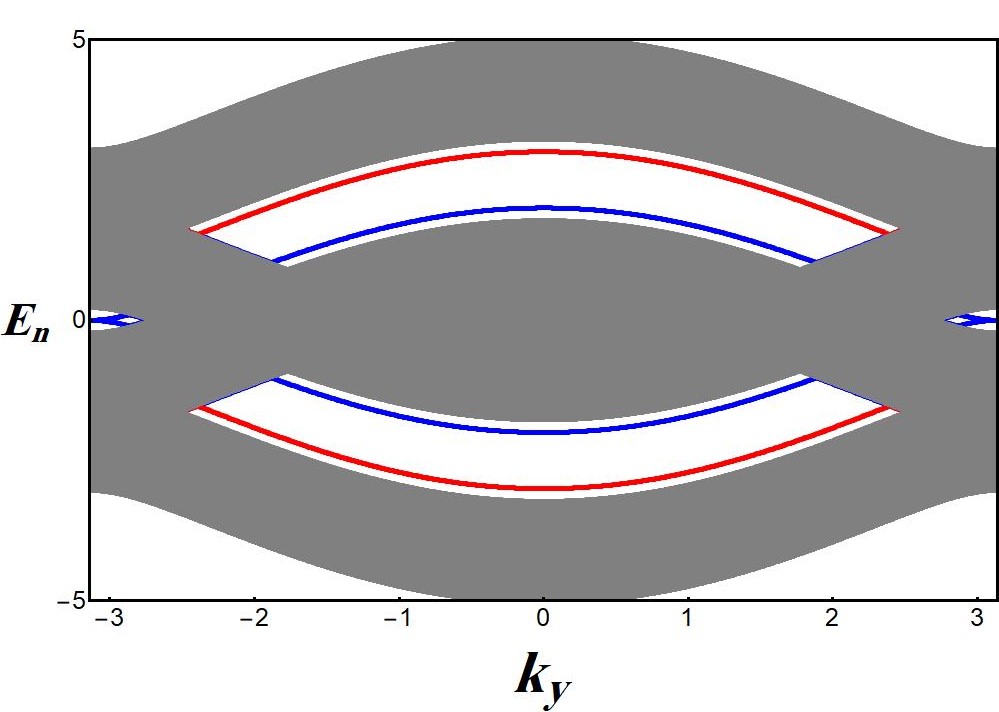}
\includegraphics[width=0.33\linewidth]{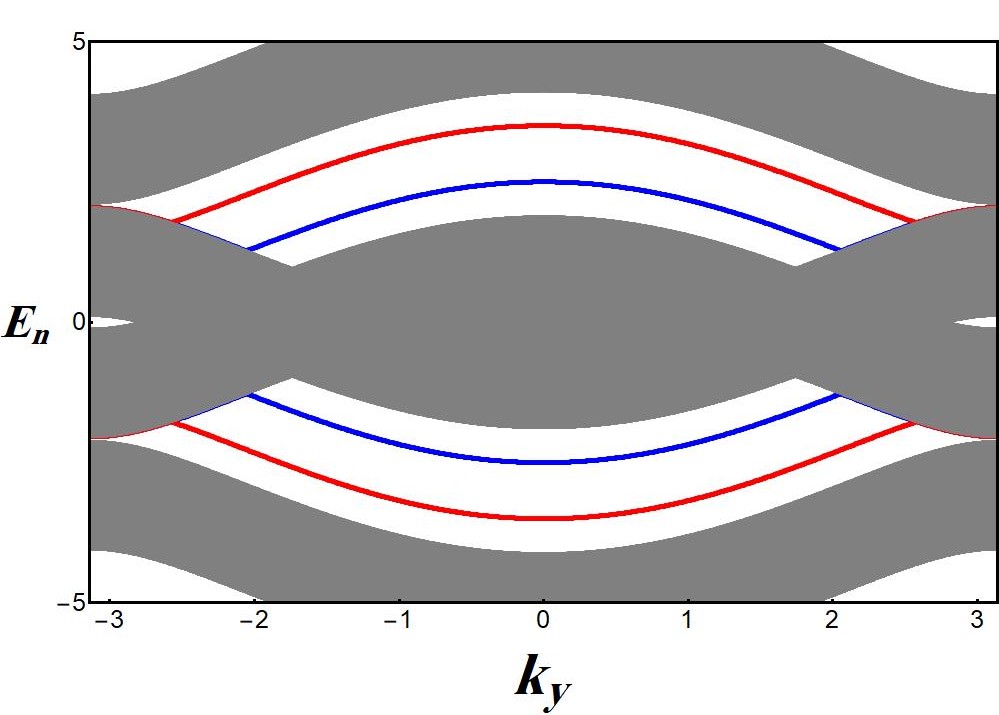}}
\caption{(Color online) Spectrum of SSH ribbon along x direction for $\mu=0.5$, $w=1$ and $N_x=225$,
(a) $v=1.2$, (b) $v=1.5$, and $v=2$. As seen double degenerate edge states appear when
$|w-v|\geqslant\mu$.}
\label{figS3}
\end{figure*}
%%%%%%%%%%%%%%%%%%%%%%%%%%%%%%%%%%%%%%%%%%%%%%%%%%%%%%%
%
Moreover, we will introduce the basis in which the eigenstates of $H_{\textbf{k}}$ are obtained as linear combinations of even-parity fermion states. Here, we present the link between the two approaches.
As $H_{\textbf{k}}$ in Eq. (\ref{eqS2}) conserves parity (even or odd number of fermions), it is sufficient to consider the even-parity subspace of the Hilbert space. This subspace is spanned by eight basis vectors,
%
%%%%%%%%%%%%%%%%%%%%%%%%%%%%%%% Eq. S4 %%%%%%%%%%%%%%%%%%%%%%%%%%%%%%%%
\begin{align}
\label{eqS4}
\no|\varphi_{1,\textbf{k}}\rangle&=|0\rangle, &  |\varphi_{2,\textbf{k}}\rangle&=c_{\textbf{k}}^{A\dag}c_{\textbf{k}}^{B\dag}|0\rangle,\\
 \no |\varphi_{3,\textbf{k}}\rangle&=c_{\textbf{k}}^{A\dag}c_{\textbf{k}}^{C\dag}|0\rangle, &
|\varphi_{4,\textbf{k}}\rangle&=c_{\textbf{k}}^{A\dag}c_{\textbf{k}}^{D\dag}|0\rangle, \\
\no|\varphi_{5,\textbf{k}}\rangle&=c_{\textbf{k}}^{B\dag}c_{\textbf{k}}^{C\dag}|0\rangle, &  |\varphi_{6,\textbf{k}}\rangle&=c_{\textbf{k}}^{B\dag}c_{\textbf{k}}^{D\dag}|0\rangle, \\
|\varphi_{7,\textbf{k}}\rangle&=c_{\textbf{k}}^{C\dag}c_{\textbf{k}}^{D\dag}|0\rangle, &
|\varphi_{8,\textbf{k}}\rangle&=c_{\textbf{k}}^{A\dag}c_{\textbf{k}}^{B\dag}c_{\textbf{k}}^{C\dag}c_{\textbf{k}}^{D\dag}|0\rangle.
\end{align}
%%%%%%%%%%%%%%%%%%%%%%%%%%%%%%%%%%%%%%%%%%%%%%%%%%%%%%%%%%%%%%%%%%%%%%%%
%
If the Hamiltonian $H_{\textbf{k}}$ of Eq. (\ref{eqS2}) is written in the basis introduced in Eq. (\ref{eqS4}),
%in the following form,
%\begin{eqnarray}
 %H_{\textbf{k}} &=&
%\begin{pmatrix}
 %0& 0& 0& 0& 0& 0& 0& 0\\
% 0& 0& w+ve^{ik_y}& 0& 0& -w-ve^{ik_y}& 0& 0\\
 %0& w+ve^{-ik_y}& 2\mu & w+ve^{ik_x}& w+ve^{-ik_x}& 0& -w-ve^{ik_y}& 0\\
 %0& 0& w+ve^{-ik_x}& 0& 0& w+ve^{-ik_x}& 0& 0\\
 %0& 0& w+ve^{ik_x}& 0& 0& w+ve^{ik_x}& 0& 0\\
 %0& -w-ve^{-ik_y}& 0& w+ve^{ik_x}& w+ve^{-ik_x}& -2\mu & w+ve^{ik_y}& 0\\
% 0& 0& -w-ve^{-ik_y}& 0& 0& w+ve^{-ik_y}& 0& 0\\
 %0& 0& 0& 0& 0& 0& 0& 0
%\end{pmatrix}.
%\nonumber\\
%\label{eqS5}
%\end{eqnarray}
the eigenstates $|\psi_{m,\textbf{k}}\rangle$ of $H_{\textbf{k}}$ in this basis can be written as
\begin{equation}
\label{eqS6}|\psi_{m,\textbf{k}}\rangle
=\sum_{j=1}^{8}p_{m,\textbf{k}}^{j}|\varphi_{j,\textbf{k}}\rangle,
\end{equation}
where $|\psi_{m,\textbf{k}}\rangle$ is a non-normalized eigenstate of $H_{\textbf{k}}$
with the corresponding eigenvalue $\epsilon_{m,\textbf{k}}\, (m=0,\cdots,7)$, and
$p_{m,\textbf{k}}^{j}\, (j=1,\cdots,8)$ are functions of the hopping amplitudes ($w,v$), chemical potential $\mu$, and the momentum $k$.
Four eigenstates are degenerate with zero eigenvalues ($\epsilon_{2,\textbf{k}}=\epsilon_{3,\textbf{k}}=\epsilon_{4,\textbf{k}}=\epsilon_{5,\textbf{k}}=0$), with the ground state and the first excited state having negative energies
$\epsilon_{0,\textbf{k}}=-\epsilon_{7,\textbf{k}}=(\varepsilon_{1,\textbf{k}}+\varepsilon_{2,\textbf{k}}), \epsilon_{1,\textbf{k}}=-\epsilon_{6,\textbf{k}}=(\varepsilon_{1,\textbf{k}}-\varepsilon_{2,\textbf{k}})$, respectively. Here $\varepsilon_{1,\textbf{k}}$ and $\varepsilon_{2,\textbf{k}}$ are the quasiparticle energies defined in Eq. (\ref{eq3}).

Each eigenstate of $H_{\textbf{k}}$ can be linked to a state in the Bogoliubov transformation formalism via their eigenvalues.
For instance, the ground state $|\psi_{0,\textbf{k}}\rangle$ of $H_{\textbf{k}}$ is characterized by the Bogoliubov mode with the corresponding negative energy quasiparticle states filled, i.e. $|\psi_{0,\textbf{k}}\rangle
= \beta^{2\dagger}_{\textbf{k}}\beta^{1\dagger}_{\textbf{k}}|V_{\textbf{k}}\rangle$, where $|V_{\textbf{k}}\rangle$ is the $k^{\text th}$ single-fermion mode of the Bogoliubov vacuum. The relation between quasiparticle operators and fermion operators is fixed by the Bogoliubov  transformation, we can calculate the Bogoliubov vacuum $|V\rangle = \bigotimes_{\textbf{k}} |V_{\textbf{k}}\rangle$ in terms of the eigenstates of $H_{\textbf{k}}$:
\begin{equation}
\label{eqS7}|V\rangle=\bigotimes_{\textbf{k}}\big(\sum_{j=1}^{8}q_{V,\textbf{k}}^{j}|\varphi_{j,\textbf{k}}\rangle\big),
\end{equation}
where $q_{V,\textbf{k}}^{j}$ are functions of the parameters $w,v$, and $\mu$, and the momentum $k$. The resulting exact expression is rather massive and unwieldy.
Let us mention that although Bogoliubov transformation formalism is very convenient for obtaining energy eigenvalues, the fermionic even (odd)-parity basis is preferable
for obtaining ground state and computing matrix elements of the time-evolved states, such as those which enter the Loschmidt overlap (LO) and Loschmidt echo (LE).
It is remarkable to mention that at the critical line $\mu=0$, the wave functions are polarized in $x$ or $y$ direction with eigenvalues
$\epsilon_{0,\textbf{k}}=-2\sqrt{v^{2}+w^{2}+2vw\cos(k_{x})}$, and $\epsilon_{1,\textbf{k}}=-2\sqrt{v^{2}+w^{2}+2vw\cos(k_{y})}$.
\vskip 1.0 cm

\section{Topological Phase Transition, Zak Phase and symmetries\label{ap-B}}
In the main text we discussed that, generally 2D SSH model in the presence of staggered chemical potential is always gapped, except at $\mu=0$, where gap closing occurs at $|k_{x}|=|k_{y}|$ for arbitrary values of $v/w$. Therefore, quantum phase transition occurs at $\mu=0$ between two distinct
insulator phases. Moreover, in the absence of chemical potential, 2D SSH model exhibits a topological phase transition at $v=w$ from a trivial topological phase $v<w$ to a nontrivial topological phase $v>w$. The energy band structure of SSH ribbon with open boundary condition has been depicted in Fig. \ref{figS2} for $N_x=225$ and for different values of $v/w$ at $\mu=0$. We assumed that the lattice is translationally invariant only along the $y$ direction and is finite in the $x$ direction with $N_x$ width.
The presence of a nontrivial topological phase is confirmed by the existence of double degenerate edge states in energy spectrum for $v/w>1$ \cite{Liu2017,Obana2019}. In addition, the study of vectored Zak phase yields a fractional wave polarization in each direction \cite{Liu2017,Obana2019}.

We have also calculated the vectored Zak phase in the 2D SSH model in the presence of the staggered chemical potential as follows \cite{Huang2014,Resta1994}
%
%%%%%%%%%%%%%%%%%%%%%%%%%%%%%%% Eq. S10 %%%%%%%%%%%%%%%%%%%%%%%%%%%%%%%%
\begin{eqnarray}
\mathcal{Z}_l=-i \int_0^{2\pi}\langle u_j(k)|\frac{\partial}{\partial k_l}|u_j(k)\rangle dk_l,\;\; l=x,y,
\label{eqS8}
\end{eqnarray}
%%%%%%%%%%%%%%%%%%%%%%%%%%%%%%%%%%%%%%%%%%%%%%%%%%%%%%%%%%%%%%%%%%%%%%%
%
where $|u_j(k)\rangle$ is the eigenvector of the Bloch Hamiltonian given in Eq. (\ref{eqS3}) corresponding to occupied (negative) energy band $\varepsilon_{j, \textbf{k}}$ \cite{Liu2017,Obana2019,Huang2014,Resta1994}.
%%
%%%%%%%%%%%%%%%%%%%%%%%%  Fig.S4   %%%%%%%%%%%%%%%%%%%%%%%
%\begin{figure}[h]
%\includegraphics[width=0.5\columnwidth]{zakphase.pdf}
%\caption{(Color online) Density plot of 2D Zak phase of extended SSH model versus $v/w$ and $\mu/w$.
%The 2D Zak Phase has a finite value $\mathcal{Z}_x=\mathcal{Z}_y=\pi$ for $v>w$,
%and is zero $\mathcal{Z}_x=\mathcal{Z}_y=0$ for $v<w$ in the absence/presence of the staggered chemical potential $\mu$.}
%\label{figS4}
%\end{figure}
%%%%%%%%%%%%%%%%%%%%%%%%%%%%%%%%%%%%%%%%%%%%%%%%%%%%%%%%
%%
Fig. \ref{fig2}(b) represents the density plot of 2D Zak phase $(Z_x,Z_y)$ versus $\mu/w$ and $v/w$.
Our results show a finite Zak phase $(Z_x,Z_y)=(\pi,\pi)$ for $v/w>1$, accompanying the fractional wave polarization $(P_x,P_y)=(1/2,1/2)$ \cite{Liu2017,Obana2019} and a zero $(Z_x,Z_y)=(0,0)$ Zak phase for the topologically trivial phase ($v/w<1$), for any value of $\mu$.
Although, the 2D Zak phase shows that there still exist topological phase transition at $v/w=1$, in the presence of the staggered chemical potential,
the edge states are not robust against perturbation for $\mu>|w-v|$ (Fig. \ref{figS3}).

Therefore, no quantum phase transition occurs at $v=w$, in the presence of chemical potential. In other words, the 2D Zak phase is not a proper indicator to describe the topological phase transition in 2D SSH model in the presence of staggered chemical potential.
Furthermore, by calculation of Chern number in continuum and discretized Brillouin Zone \cite{asboth2016,Takahiro2005} of the 2D extended SSH model, we obtain that Chern number is zero in the presence of staggered chemical potential for all values of hopping amplitudes.

The 2D SSH Hamiltonian has different topological classes in the absence/presence of the staggered chemical potential $\mu$.
In the absence of staggered chemical potential, the model has time-reversal symmetry, $TH(\textbf{k})T^{-1}=H(-\textbf{k})$, where, the time-reversal operator is defined by the complex conjugate operator $T=\mathcal{K}$. Moreover, it has the chiral symmetry, which is defined by $CH(\textbf{k})C^{-1}=-H(\textbf{k})$ where the sublattice symmetry operator for 2D SSH model is given by
%
%%%%%%%%%%%%%%%%%%%%%%%%%%%%%%% Eq. S11 %%%%%%%%%%%%%%%%%%%%%%%%%%%%%%%%
\begin{eqnarray}
\label{eqS9} C &=&
\begin{pmatrix}
 1&0& 0& 0\\
 0& -1& 0& 0\\
 0& 0& 1& 0\\
 0& 0& 0& -1
\end{pmatrix}
\end{eqnarray}
%%%%%%%%%%%%%%%%%%%%%%%%%%%%%%%%%%%%%%%%%%%%%%%%%%%%%%%%%%%%%%%%%%%%%%%
%
Furthermore, we observe that the Hamiltonian satisfies $PH(\textbf{k})P^{-1}=-H(-\textbf{k})$, where $P=C T$. This means the system has also particle hole symmetry.
Since $T^{2}$, $C^{2}=1$, and $P^{2}=1$, the topological class of 2D SSH model in the absence of chemical potential is BDI \cite{Chiu2016}.

In the presence of chemical potential $\mu\neq0$, inspection of the Hamiltonian of Eq. (\ref{eqS3}) reveals that only time-reversal symmetry of the model is preserved.
The breakdown of chiral and particle-hole symmetries by the staggerd chemical potential, changes the topological class of 2D SSH model to be AI.
There is no gap closing and no symmetry breaking at $v/w=1$, in the presence of the staggered chemical potential.
Hence, the changes in the 2D Zak phase at $v/w=1$ and $\mu\neq0$ does not correspond to a real topological phase transition. So, the topological phase transition occurs only in the absence of chemical potential at $v/w=1$.
It is worthwhile to mention that the line $\mu=0$, is the phase boundary between two insulator phases.
%%%%%%%%%%%%%%%%%%%%%%%  Fig.S2   %%%%%%%%%%%%%%%%%%%%%%%
\begin{figure*}
\centerline{\includegraphics[width=0.33\linewidth]{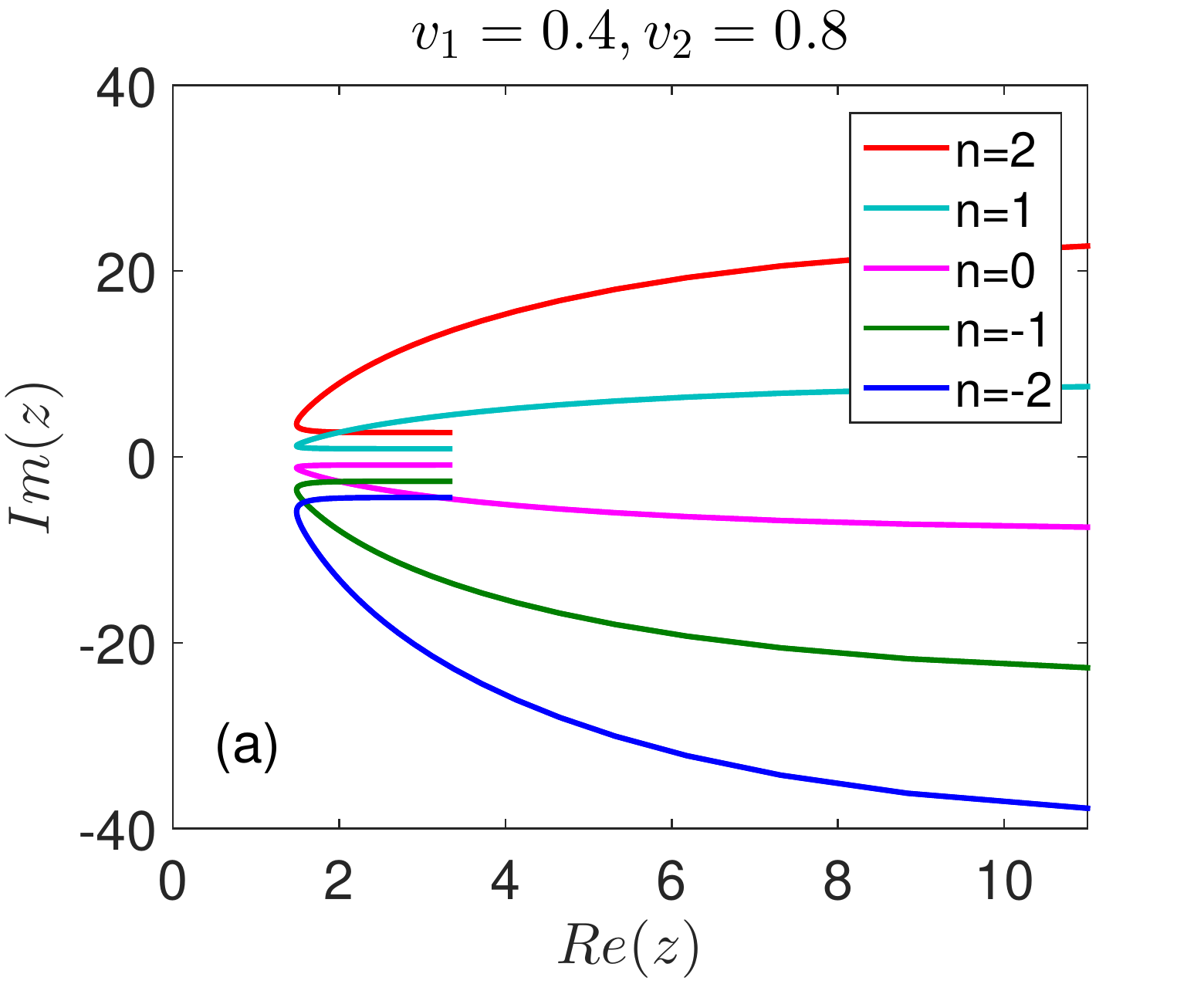}
\includegraphics[width=0.33\linewidth]{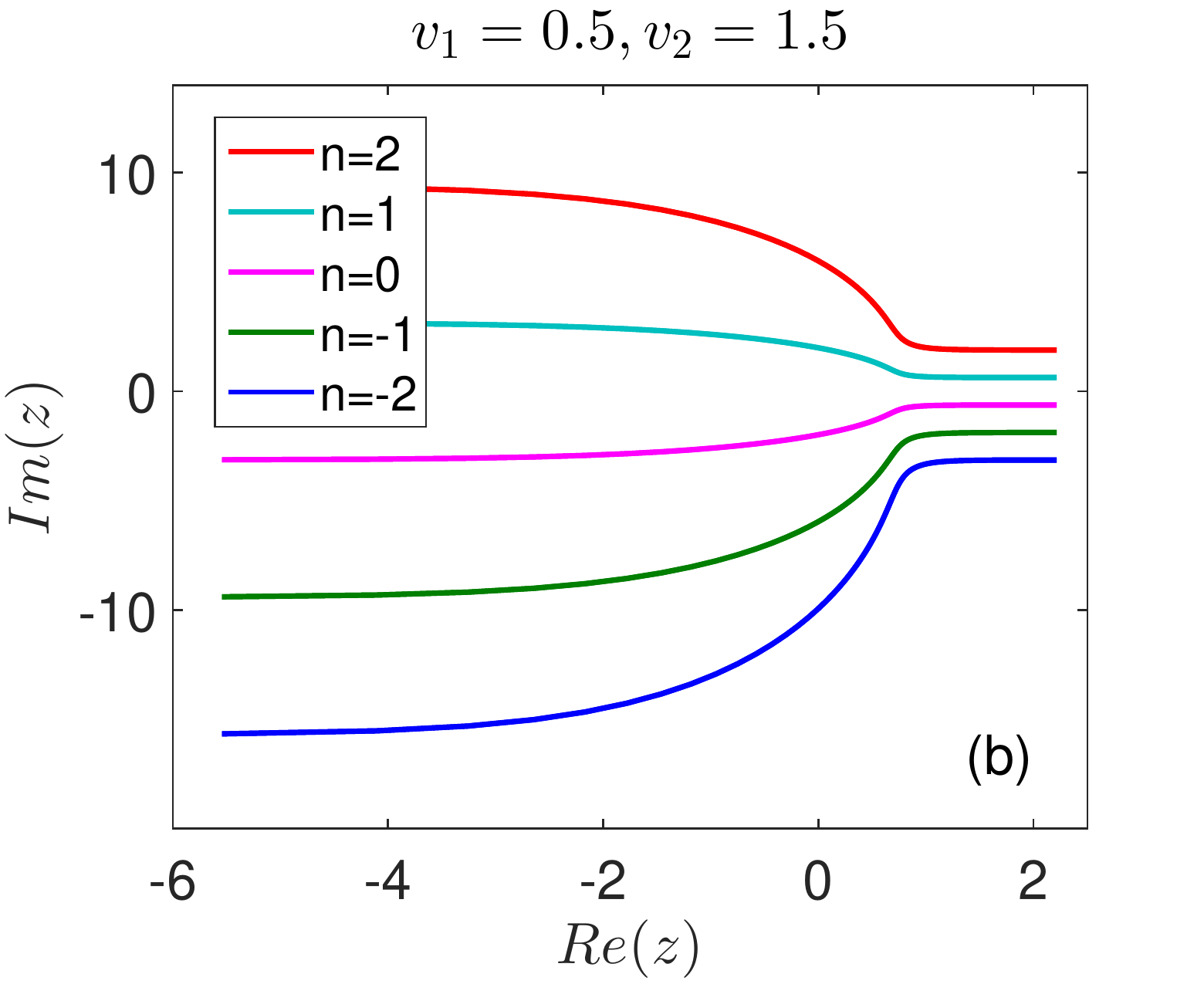}
\includegraphics[width=0.33\linewidth]{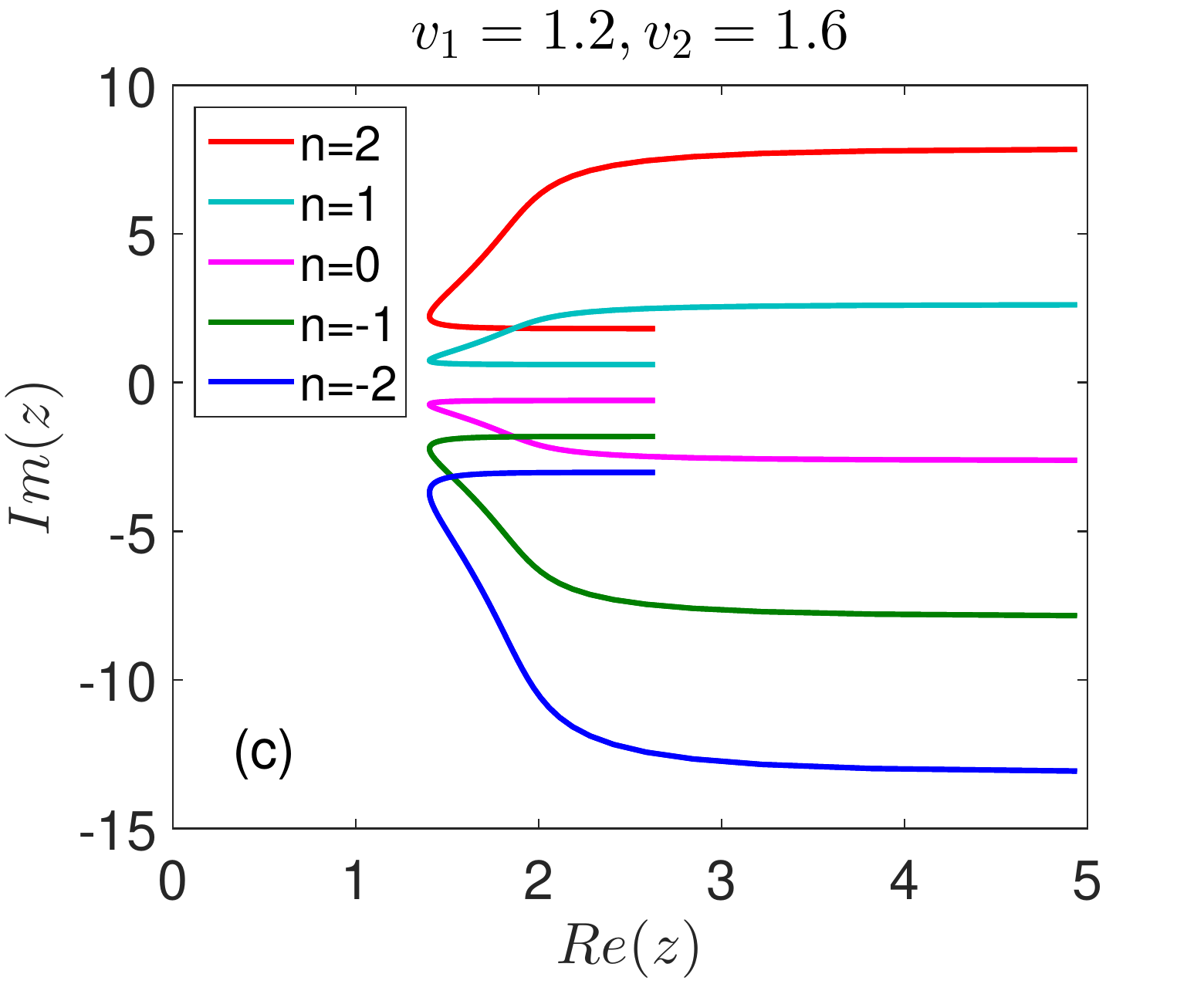}}
\caption{(Color online) Lines of Fisher zeros at critical line $\mu=0$,
for quenches (a) from $v_1=0.4$ to $v_2=0.8$ (b), from $v_1=0.5$ to $v_2=1.5$
(c) from $v_1=1.2$ to $v_2=1.6$. As seen, for quenches across the topological
quantum phase transition point $v/w=1$ i.e. from $v_1=0.5$ to $v_2=1.5$,
the lines of Fisher zeros cross the real-time axis which interpreted as a
dynamical phase transition.}
\label{figS4}
\end{figure*}
%%%%%%%%%%%%%%%%%%%%%%%%%%%%%%%%%%%%%%%%%%%%%%%%%%%%%%%%%%%%%%%%%%%%%%%
%
%
%%%%%%%%%%%%%%%%%%%%%%%  Fig.S1   %%%%%%%%%%%%%%%%%%%%%%%
\begin{figure*}
\centerline{\includegraphics[width=0.4\linewidth]{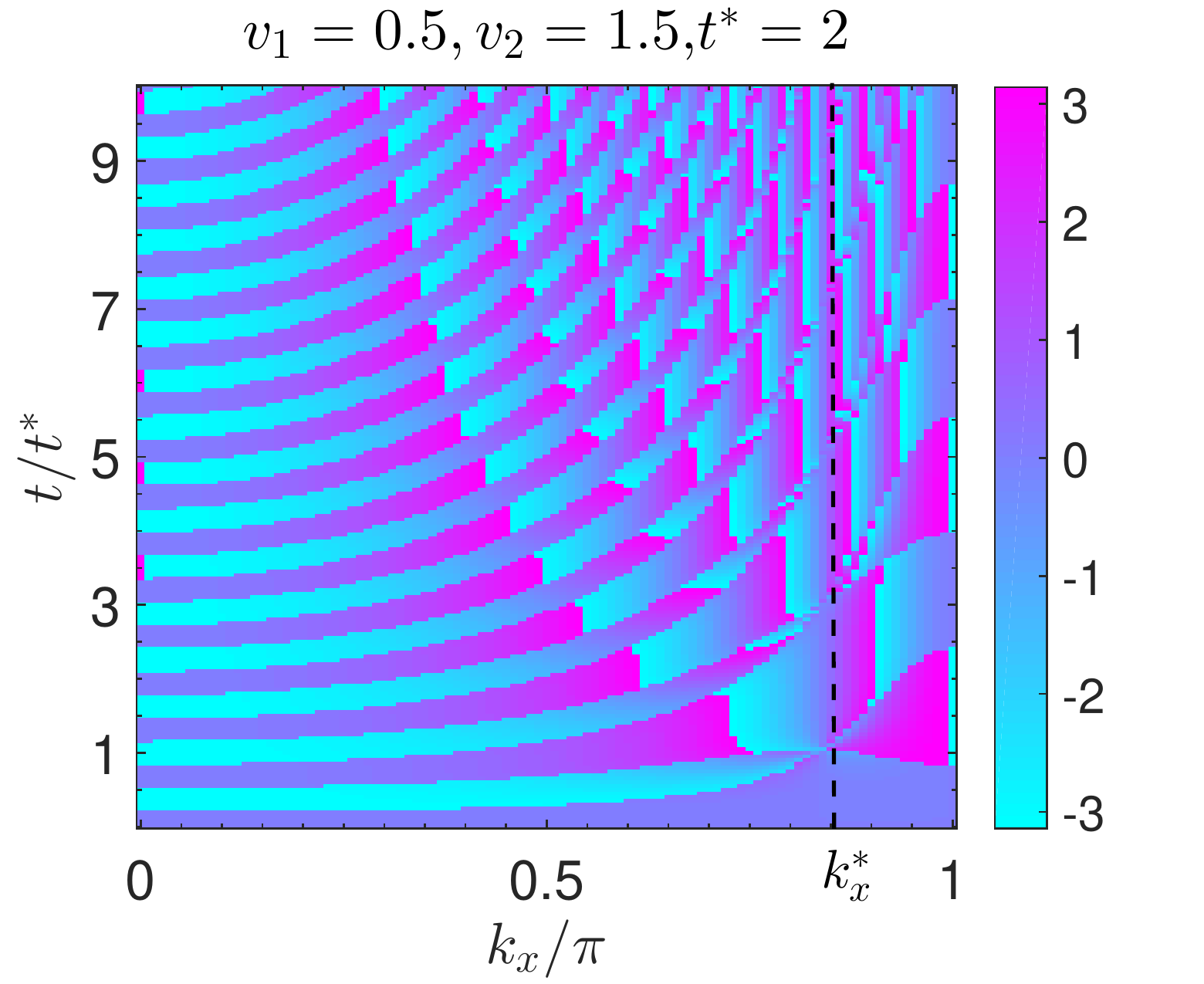}
\includegraphics[width=0.4\linewidth]{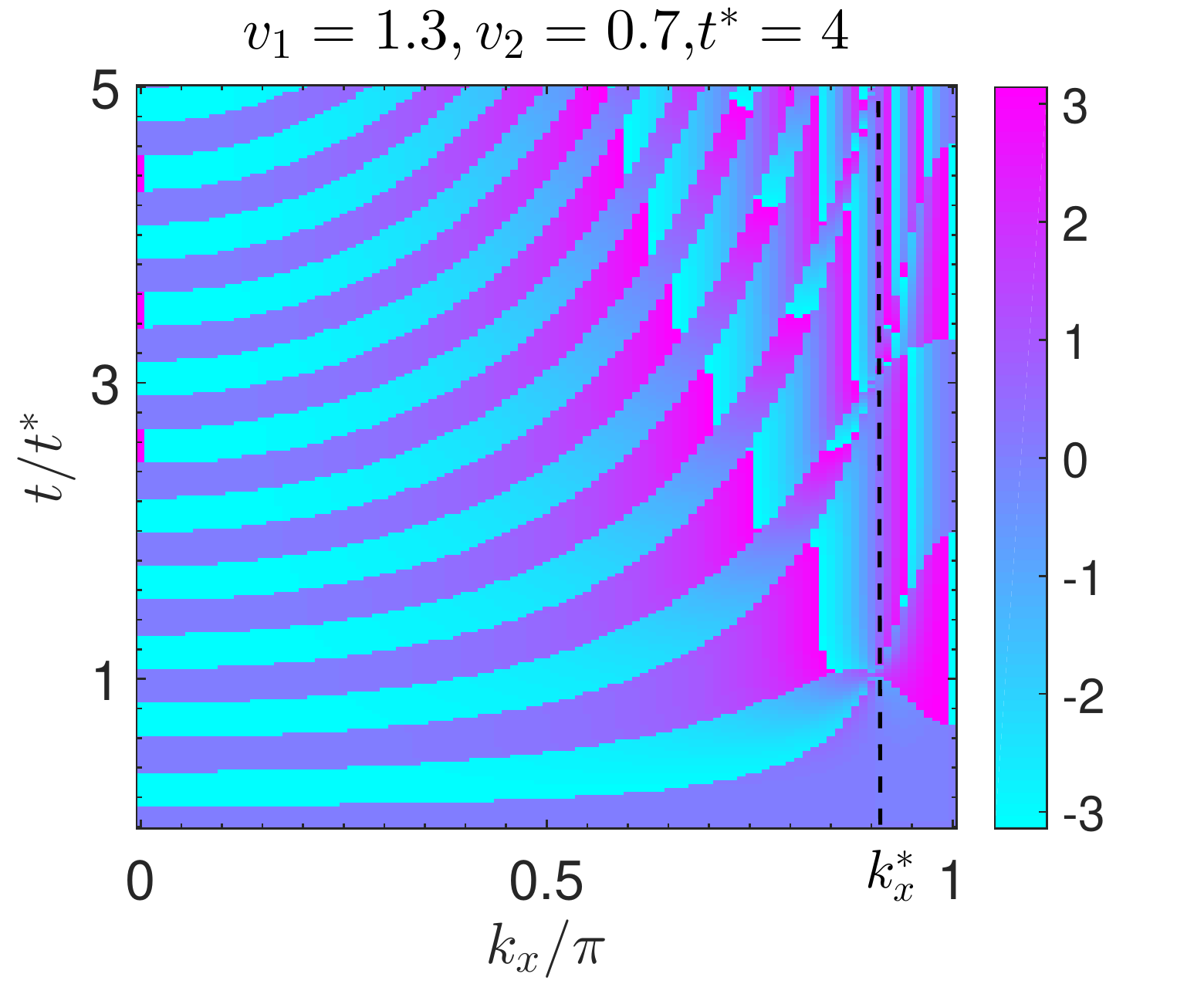}}
\caption{(Color online) Density plots of the Pancharatnam geometric phase
$\phi^G(\textbf{k},t)$ as a function of lattice momentum and time,
at critical line $\mu=0$, for two different quenches of inter-plaquette
hopping amplitude $v_1\rightarrow v_2$ across the topological phase
transition point $v/w=1$, correspond to Fig. \ref{fig3}(b).
Time is measured in units of the critical time $t^{\ast}$ and the critical
wave vector $k_x^{\ast}$ at which nonanalyticities occur is marked with a black dotted line.}
\label{figS5}
\end{figure*}
%%%%%%%%%%%%%%%%%%%%%%%%%%%%%%%%%%%%%%%%%%%%%%%%%%%%%%%%%%%%%%%%%%%%%%%
%
%
\section{Loschmidt overlap and Loschmidt echo \label{ap-C}}
By using eigenvalues and eigenvectors of $H_{\textbf{k}}$ in the basis introduced in Eq. (\ref{eqS4}), we can calculate Loschmidt overlap for a quench from initial Hamiltonian $\mathcal{H}(v_1,\mu_1)$ to final Hamiltonian $\mathcal{H}(v_2,\mu_2)$. Note that, we will not change the intracell hopping parameter $w$ in the quenching process (we set $w=1$ as the energy scale
in our numerical simulations).
If the initial state is prepared to be the ground state of the initial Hamiltonian $|\psi_{0}(v_1,\mu_1)\rangle$, by using $\mathcal{H}=\sum_{\textbf{k}}H_{\textbf{k}}$, the mode decomposition of the Loschmidt overlap ${\cal L}({v_1},{v_2},\mu_{1},\mu_{2},t)$, in the general case of 2D SSH model in the presence of staggered chemical potential takes the form,

%
%%%%%%%%%%%%%%%%%%%%%%%%%%%%%%% Eq. S8 %%%%%%%%%%%%%%%%%%%%%%%%%%%%%%%%
\bea
\label{eqS10}
\no &&{\cal L}({v_1},{v_2},\mu_{1},\mu_{2},t)= \prod_{\textbf{k}}{\cal L}_\textbf{k}({v_1},{v_2},\mu_{1},\mu_{2},t),\\
\no &&{\cal L}_\textbf{k}({v_1},{v_2},\mu_{1},\mu_{2},t)=\\
\no &&\frac{1}{N_{0,\textbf{k}}(v_{1},\mu_{1})} \langle\psi_{0,\textbf{k}}({v_1},\mu_{1})|
e^{-iH(v_{2},\mu_{2})t}|\psi_{0,\textbf{k}}(v_{1},\mu_{1})\rangle=\\
\no
&&\frac{1}{N_{0,\textbf{k}}(v_{1},\mu_{1})}
\sum_{m=0}^{7}\frac{e^{-i \epsilon_{m,\textbf{k}}(v_{2},\mu_{2})t}}{N_{m,\textbf{k}}(v_{2},\mu_{2})}|
\langle\psi_{m,\textbf{k}}(v_{2},\mu_{2})|\psi_{0,\textbf{k}}(v_{1},\mu_{1})\rangle|^{2}=\\
&&\sum_{m=0}^{7} e^{-i \epsilon_{m,\textbf{k}}(v_{2},\mu_{2})t}F_{m,\textbf{k}}({v_1},{v_2},\mu_{1},\mu_{2}),
\eea
%%%%%%%%%%%%%%%%%%%%%%%%%%%%%%%%%%%%%%%%%%%%%%%%%%%%%%%%%%%%%%%%%%%%%%%
%
where $N_{m,\textbf{k}}(v,\mu)=|\langle\psi_{m,\textbf{k}}(v,\mu)|\psi_{m,\textbf{k}}(v,\mu)\rangle|$
is the normalization factor of the eigenstate $|\psi_{m,\textbf{k}}(v,\mu)\rangle$, and $F_{m,\textbf{k}}({v_1},{v_2},\mu_{1},\mu_{2})=
\Big|\langle\psi_{m,\textbf{k}}({v_2},\mu_{2})
|\psi_{0,\textbf{k}}(v_{1},\mu_{1})\rangle\Big|^{2}/\Big[N_{0,\textbf{k}}(v_{1},\mu_{1})N_{m,\textbf{k}}(v_{2},\mu_{2})\Big]$.\\

For a quench at the critical line $\mu_{1}=\mu_{2}=0$ from $v_{1}$ to $v_{2}$ the LO reduces to the simple form
%
%%%%%%%%%%%%%%%%%%%%%%%%%%%%%%% Eq. %%%%%%%%%%%%%%%%%%%%%%%%%%%%%%%%
\bea
&&{\cal L}({v_1},{v_2},t)=\prod_{\textbf{k}}\Big[e^{-i \epsilon_{0,\textbf{k}}(v_{2})t}F_{0,\textbf{k}}({v_1},{v_2})
\no \\
&&+F_{2,\textbf{k}}({v_1},{v_2})+e^{i \epsilon_{0,\textbf{k}}(v_{2})t}F_{7,\textbf{k}}({v_1},{v_2})\Big].
\eea
%%%%%%%%%%%%%%%%%%%%%%%%%%%%%%%%%%%%%%%%%%%%%%%%%%%%%%%%%%%%%%%%%%%%%%%
%
The modulus of the LO, Loschmidt echo (LE), at the critical line $\mu=0$ is given by Eq. (\ref{eq5}), where
$A_{\textbf{k}}=4F_{0,\textbf{k}}({v_1},{v_2})F_{7,\textbf{k}}({v_1},{v_2})$ and $B_{\textbf{k}}=4F_{2,\textbf{k}}({v_1},{v_2})\Big[F_{0,\textbf{k}}({v_1},{v_2})+F_{7,\textbf{k}}({v_1},{v_2})\Big]$.\\

In the presence of staggered chemical potential ($\mu\neq0$), the LO of the model is generally given by,
%
%%%%%%%%%%%%%%%%%%%%%%%%%%%%%%% Eq.  %%%%%%%%%%%%%%%%%%%%%%%%%%%%%%%%
\bea
\no
{\cal L}({v_1},{v_2},\mu_1,\mu_2,t)&=&\prod_{\textbf{k}}\Big[e^{-i \epsilon_{0,\textbf{k}}(v_{2},\mu_2)t}F_{0,\textbf{k}}({v_1},{v_2},\mu_1,\mu_2)\\
\no &&+e^{-i \epsilon_{1,\textbf{k}}(v_{2},\mu_2)t}F_{1,\textbf{k}}({v_1},{v_2},\mu_1,\mu_2)\\
\no &&+F_{2,\textbf{k}}({v_1},{v_2},\mu_1,\mu_2)\\
 &&+F_{3,\textbf{k}}({v_1},{v_2},\mu_1,\mu_2)\\
\no &&+e^{i \epsilon_{1,\textbf{k}}(v_{2},\mu_2)t}F_{6,\textbf{k}}({v_1},{v_2},\mu_1,\mu_2)\\
\no &&+e^{i \epsilon_{0,\textbf{k}}(v_{2},\mu_2)t}F_{7,\textbf{k}}({v_1},{v_2},\mu_1,\mu_2)\Big].
\eea
%%%%%%%%%%%%%%%%%%%%%%%%%%%%%%%%%%%%%%%%%%%%%%%%%%%%%%%%%%%%%%%%%%%%%%%
%
In that case, LE is obtained as given in Eq. (\ref{eq6}), where
%
%%%%%%%%%%%%%%%%%%%%%%%%%%%%%%% Eq. %%%%%%%%%%%%%%%%%%%%%%%%%%%%%%%%
\bea
\no
A_{0,\textbf{k}}&=&4F_{0,\textbf{k}}({v_1},{v_2},\mu_1,\mu_2)F_{7,\textbf{k}}({v_1},{v_2},\mu_1,\mu_2),\\
\no
A_{1,\textbf{k}}&=&4\Big[F_{2,\textbf{k}}({v_1},{v_2},\mu_1,\mu_2)+F_{3,\textbf{k}}({v_1},{v_2},\mu_1,\mu_2)\Big]\\
\no &&\times\Big[F_{0,\textbf{k}}({v_1},{v_2},\mu_1,\mu_2)+F_{7,\textbf{k}}({v_1},{v_2},\mu_1,\mu_2)\Big],\\
\no
A_{2,\textbf{k}}&=&4F_{1,\textbf{k}}({v_1},{v_2},\mu_1,\mu_2)F_{6,\textbf{k}}({v_1},{v_2},\mu_1,\mu_2),\\
\no
A_{3,\textbf{k}}&=&4\Big[F_{2,\textbf{k}}({v_1},{v_2},\mu_1,\mu_2)+F_{3,\textbf{k}}({v_1},{v_2},\mu_1,\mu_2)\Big]\\
\no &&\times\Big[F_{1,\textbf{k}}({v_1},{v_2},\mu_1,\mu_2)+F_{6,\textbf{k}}({v_1},{v_2},\mu_1,\mu_2)\Big],\\
\no
A_{4,\textbf{k}}&=&4\Big[F_{0,\textbf{k}}({v_1},{v_2},\mu_1,\mu_2)F_{6,\textbf{k}}({v_1},{v_2},\mu_1,\mu_2)\\
\no &&+F_{1,\textbf{k}}({v_1},{v_2},\mu_1,\mu_2)F_{7,\textbf{k}}({v_1},{v_2},\mu_1,\mu_2)\Big],\\
\no
A_{5,\textbf{k}}&=&4\Big[F_{0,\textbf{k}}({v_1},{v_2},\mu_1,\mu_2)F_{1,\textbf{k}}({v_1},{v_2},\mu_1,\mu_2)\\
 &&+F_{6,\textbf{k}}({v_1},{v_2},\mu_1,\mu_2)F_{7,\textbf{k}}({v_1},{v_2},\mu_1,\mu_2)\Big].
\eea
%%%%%%%%%%%%%%%%%%%%%%%%%%%%%%%%%%%%%%%%%%%%%%%%%%%%%%%%%%%%%%%%%%%%%%%
%

\vskip 1.5 cm
\section{Fisher zeroes, the Pancharatnam geometrical phase and topological invariant \label{ap-D}}
From a formal point of view, the Loschmidt overlap ${\cal L}(z)$,
has the form of a partition function with boundaries fixed by the initial state ${\cal L}(z)=\langle\Psi_0|\exp(-z\mathcal{H})|\Psi_0\rangle$
with $z \in \mathbb{C}$. Then, a general complex parameter $z$ allows to investigate the structure of ${\cal L}(z)$ in the whole complex
plane. In a spirit similar to the classical case, one then looks for zeros of ${\cal L}(z)$ known as Fisher zeros.
If such zeros close on the imaginary axis, $z_{n}\rightarrow it_{n}^{\ast}$, nonanalyticities will occur in the real time evolution
of the quantum system at times $t_{n}^{\ast}$, which is interpreted as a dynamical quantum phase transition.
The sketches of lines of Fisher zeros are represented in Fig. \ref{figS4} for different quenches of inter-plaquette hopping amplitude $v$ at critical line $\mu=0$. It reveals that for quenches within the same phase where $v_1,v_2<1$ and $v_1,v_2>1$, the lines of
Fisher zeroes do not cut the imaginary axis, while the imaginary axis is crossed by Fisher zeros lines if the system quenched through the topological quantum phase transition point.

As mentioned in the main text, a dynamical topological order parameter (DTOP) has been proposed in order to establish the connection between dynamical
quantum phase transitions (DQPTs) and the standard theory of phase transitions.
First, the LO is written as ${\cal L}_{\textbf{k}}=|{\cal L}_{\textbf{k}}|e^{i\phi_{\textbf{k}}(t)}$ where the phase $\phi_{\textbf{k}}(t)$ is the sum
of the dynamic phase $\phi^{D}_{\textbf{k}}(t)=-\int_0^t dt'\langle \psi_{0,\textbf{k}}(t')|\mathcal{H}_f|\psi_{0,\textbf{k}}(t')\rangle$, and the geometric phase (Pancharatnam geometrical phase) $\phi^G_{\textbf{k}}(t)$ \cite{Budich2016}. Now, in two dimension, the DTOP is given by \cite{Bhattacharya2017b},
%%%%%%%%%%%%%%%%%%%%%%%%%%%%%%%%% Eq. S%%%%%%%%%%%%%%%%%%%%%%%%%%%%%%%%
\begin{eqnarray}
\nu_D(t)=\frac{1}{2\pi}\int_0^\pi \mathrm{d}k_x\frac{\partial}{\partial k_x}[\int_0^\pi\mathrm{d}k_y\frac{\partial\phi^G_{\textbf{k}}(t)}{\partial k_y}].
\label{eqS11}
\end{eqnarray}
%%%%%%%%%%%%%%%%%%%%%%%%%%%%%%%%%%%%%%%%%%%%%%%%%%%%%%%%%%%%%%%%%%%%%%%
%
A quantized DTOP appears due to the existence of some critical wave vectors, at which the phase of the LO
shows a $\pi$ phase slip, originating from the change in the sign of the LO at the critical wave vector $\textbf{k}^{*}$ and critical time $t^{*}$.
Fig. \ref{figS5} shows density plots of the geometric phase as a function of $k$ and $t$ for different quenches of hopping amplitude $v$ across topological quantum phase transition point $v/w=1$ at critical line $\mu=0$, corresponding to Fig. \ref{fig3}(b). The plots display singular changes in successive critical times $t^{\ast}_n=(2n+1)t^{\ast}$ at critical momentum $k_x^{\ast}$.

\section{G. Experimental scheme\label{ap-E}}
\begin{figure}
\includegraphics[width=0.5\columnwidth]{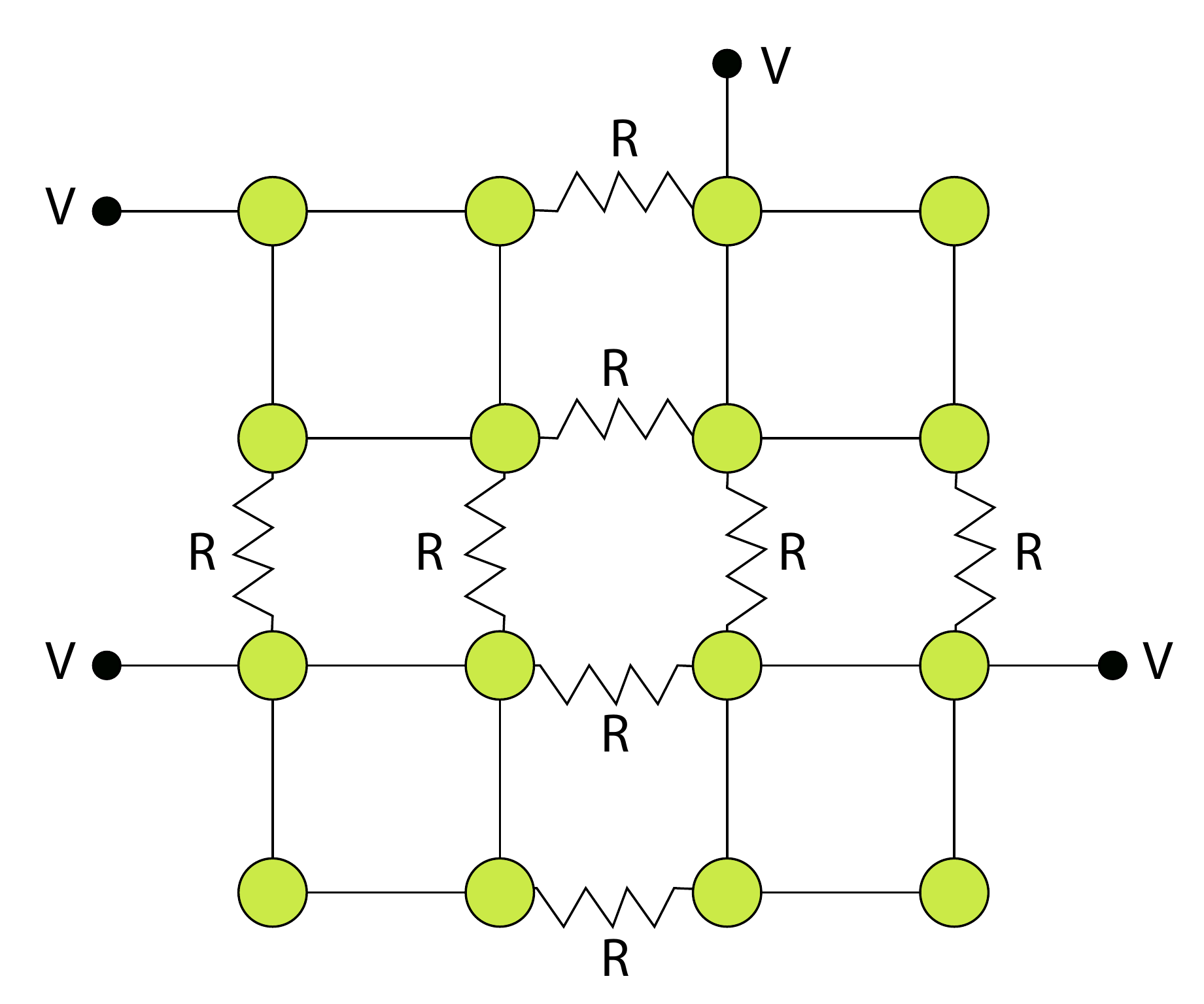}
\caption{A proposed experimental setup of 2D SSH model consists of sixteen
identical nanomechanical out-of-plane oscillators (green circles). The
unit cell of 2D SSH model is composed of four adjacent oscillators. The fundamental modes of
these oscillators are used as the hopping objects. The intra-plaquette and inter-plaquette hopping amplitudes can be realized and tuned by applying DC and AC voltages between every four adjacent oscillators.}
\label{figS6}
\end{figure}

We briefly discuss the experimental realization of
our findings on the nonequilibrium dynamics of the 2D SSH model, simulated by nanoelectromechanical resonators \cite{PhysRevB.100.024310,PhysRevB.101.174303}. We realize the 2D SSH model by a parametrically reconfigurable coupled mechanical oscillators, shown in Fig. \ref{figS6}.
The lattice consists of 16 identical oscillators. Every unit cell of 2D SSH model is composed of four oscillators on a plaquette. The intra-plaquette and inter-plaquette hopping amplitudes are controlled by different parametric couplings, which are realized by applying voltages $V_{DC}+V_{AC}cos(\omega_pt)$ on adjacent oscillators. The system can be reconfigurable by tunning voltages. By applying the voltage $V_{DC}+V_{AC}cos(\omega_pt)$ in the first oscillator, four oscillators on a plaquette couple to each other. Then, one can measure the frequency response at three other oscillators of a plaquette, where the split between two frequency peaks gives the parametric coupling strength. By tuning different couplings between adjacent oscillators, various structures of 2D SSH model can be realized to study quench dynamics. An edge state can be chosen as the initial state by the excitation of an edge oscillator with a sinusoidal wave. Then, turnning on all the coupling voltages between adjacent oscillators to configure two different structures of 2D SSH model ($v<w$ and $v>w$) as a final state. As the time evolves, one can measure the response spectrum at the edge oscillator. Therefore, DQPT can be investigated via dynamical evolution of initial edge excitation.
\vskip 1.0 cm
\bibliography{Refs}
\end{document}